\pdfoutput=1

\documentclass[%
 reprint,
 superscriptaddress,
 floatfix,
 nobibnotes,
 amsmath,amssymb,
 aps,
 showkeys,
 longbibliography,
]{revtex4-2}

\usepackage{graphicx}
\usepackage{dcolumn}
\usepackage{bm}
\usepackage{chemformula}
\usepackage[version=3]{mhchem}
\usepackage{multirow}
\usepackage[separate-uncertainty=true,separate-uncertainty,multi-part-units=single]{siunitx}
\usepackage{hyperref}
\hypersetup{colorlinks=true,allcolors=magenta}
\usepackage{array}

\begin{document}

\title{Influence of Ultramicroporosity and Surface Chemistry on Dynamic \ce{CO2} Capture in Activated Carbons}

\author{S. Gooijer}
\affiliation{Materials Simulation and Modelling, Department of Applied Physics and Science Education, Eindhoven University of Technology, PO Box 513, Eindhoven, 5600 MB, The Netherlands}
\author{P. Goosen}
\affiliation{Materials Simulation and Modelling, Department of Applied Physics and Science Education, Eindhoven University of Technology, PO Box 513, Eindhoven, 5600 MB, The Netherlands}
\author{C. G. Díaz-Maroto}
\affiliation{Thermochemical Processes Unit, IMDEA Energy Institute, Avda. Ramón de la Sagra, 3, 28935 Móstoles, Madrid, Spain}
\author{I. Moreno}
\affiliation{Thermochemical Processes Unit, IMDEA Energy Institute, Avda. Ramón de la Sagra, 3, 28935 Móstoles, Madrid, Spain}
\author{S. Calero}
\email{s.calero@tue.nl}
\affiliation{Materials Simulation and Modelling, Department of Applied Physics and Science Education, Eindhoven University of Technology, PO Box 513, Eindhoven, 5600 MB, The Netherlands}
\author{J. Fermoso}
\email{javier.fermoso@csic.es}
\affiliation{Instituto de Catálisis y Petroleoquímica (ICP), CSIC, C/ Marie Curie 2, 28049 Madrid, Spain}
\author{J.M. Vicent-Luna}
\email{j.vicent.luna@tue.nl}
\affiliation{Materials Simulation and Modelling, Department of Applied Physics and Science Education, Eindhoven University of Technology, PO Box 513, Eindhoven, 5600 MB, The Netherlands}

\date{June 16, 2026}

\begin{abstract}
Activated carbons are promising adsorbents for post-combustion \ce{CO2} capture due to their high surface area, tunable microporosity, and resistance to moisture and flue-gas impurities. Despite extensive equilibrium adsorption studies, the dynamic behavior of activated carbons under fixed-bed operating conditions relevant to post-combustion \ce{CO2}/\ce{N2} remains insufficiently understood, particularly for renewable materials. In this work, the adsorption and separation behavior of \ce{CO2}/\ce{N2} mixtures on a commercial coal-derived activated carbon (WS-480) and a biomass-based activated carbon (MSP700-A900\ce{CO2}) is comparatively evaluated by combining experimental measurements and simulations. We examine the physicochemical properties of both materials, revealing that although WS-480 exhibits a higher of porosity, MSP700-A900\ce{CO2} contains a larger fraction of ultramicropores ($<$0.7 nm) and a broader distribution of oxygen-containing functional groups. These characteristics result in higher \ce{CO2} adsorption capacities for MSP700-A900\ce{CO2} in fixed-bed breakthrough experiments conducted under varying flow rates, temperatures and \ce{CO2} concentrations. We employ atomistic activated carbon models, augmented with surface functional groups as representations of WS-480 and MSP700-A900\ce{CO2}, achieving close agreement with experimental adsorption data. The validated models are subsequently used to predict \ce{CO2}/\ce{N2} separation under equilibrium and dynamic conditions, reproducing the experimental breakthrough behavior while providing molecular-level insight into the influence of pore structure and surface chemistry on adsorption performance.
\end{abstract}


\maketitle


\section{Introduction}
\label{sec:intro}
Activated carbons (ACs) are among the most widely investigated adsorbents for \ce{CO2} capture due to their advantageous physicochemical properties, including high specific surface area, tunable pore size distribution, and chemical stability, among others. Compared to other classes of adsorbents, such as zeolites, metal-organic frameworks (MOFs), or mesoporous silicas, activated carbons exhibit superior resistance to moisture and flue-gas impurities, which is particularly beneficial for post-combustion capture applications.\cite{raganati_adsorption_2021, wen_recent_2024, gonzalez_sustainable_2013, ibrahim_review_2025} Furthermore, their adsorption performance can be tailored through the appropriate selection of precursor materials and activation conditions, enabling the development of pore structures optimized for \ce{CO2} adsorption.

The adsorption of \ce{CO2} in activated carbons is primarily governed by their pore structure, particularly the presence of narrow micropores. In this regard, ultramicropores (${<}$0.7 nm) play a dominant role in enhancing \ce{CO2} adsorption at low partial pressures due to overlapping adsorption potentials within confined pores, which significantly strengthen adsorbate–adsorbent interactions.\cite{sevilla_sustainable_2011, presser_effect_2011, kundu_comprehensive_2024} In addition to textural properties, the surface chemistry of activated carbons can also influence adsorption performance, as the presence of heteroatoms or functional groups may modify the interactions between \ce{CO2} molecules and the carbon surface. Therefore, the optimization of both pore structure and surface functionality is crucial for the development of high-performance carbon adsorbents.\cite{kundu_comprehensive_2024, jedli_activated_2024, senthilkumar_unveiling_2025}

In recent years, increasing attention has been directed toward the production of activated carbons from renewable biomass resources. Biomass-based precursors, such as agricultural residues, lignocellulosic wastes, and energy crops, represent attractive alternatives to conventional fossil-based feedstocks due to their abundance, low cost, and reduced environmental impact.\cite{plaza_development_2009, soo_advancements_2024} In addition to their environmental benefits, biomass-derived activated carbons may offer an economic advantage over conventional fossil-based counterparts, particularly when low-cost agroforestry residues are used as precursors. Although the overall production cost remains strongly dependent on factors such as activation route, energy consumption, and post-treatment requirements, the wide availability and low market value of biomass wastes make them attractive feedstocks for the development of sustainable and potentially low-cost adsorbents.\cite{jaria_overview_2022} Moreover, the conversion of biomass into porous carbons contributes to waste valorization and the development of sustainable materials. When suitable activation methods are applied, biomass-derived carbons can exhibit highly developed microporous structures comparable to those of conventional activated carbons, making them promising candidates for \ce{CO2} capture applications.\cite{kundu_comprehensive_2024, plaza_development_2009, al-ghurabi_optimizing_2025} Despite the considerable research on \ce{CO2} adsorption using activated carbons, several challenges remain in understanding adsorption behavior under process conditions relevant to post-combustion \ce{CO2}/\ce{N2} separation. 

While equilibrium adsorption measurements are extensively reported in the literature, studies addressing dynamic adsorption behavior in fixed-bed systems with multicomponent gas mixtures remain comparatively limited, despite their greater relevance to industrial applications.\cite{kundu_comprehensive_2024, karimi_carbon_2023} Most available works focus on single-component isotherms and their fitting to equilibrium models, whereas fewer studies systematically investigate breakthrough behavior under varying operating conditions such as temperature, flow rate, and feed composition.\cite{karimi_carbon_2023} Moreover, the integration of experimental fixed-bed data with rigorous modeling approaches is still relatively scarce, particularly for biomass-derived activated carbons.\cite{jedli_activated_2024} This lack of comprehensive experimental-modeling studies hinders the accurate prediction of adsorption performance and limits the development of reliable design and scale-up methodologies for adsorption-based \ce{CO2} capture processes.\cite{karimi_carbon_2023, sircar_basic_2006} This is particularly critical when transitioning from laboratory-scale experiments to industrial applications, where mass transfer limitations and non-ideal flow behavior play a decisive role. Therefore, further research combining dynamic experiments with validated theoretical models is required to bridge this gap and to better understand the performance of activated carbons under operating conditions relevant to post-combustion \ce{CO2}/\ce{N2} separation.

In this work, the \ce{CO2} adsorption performance of two activated carbons with different origins will be comparatively evaluated: a commercial coal-derived activated carbon and a renewable activated carbon. The renewable adsorbent is produced via physical activation with \ce{CO2} at 900 ºC using a biochar precursor obtained from the pyrolysis of miscanthus straw pellets. Dynamic adsorption experiments are carried out in a fixed-bed reactor using \ce{CO2}/\ce{N2} mixtures under different operating conditions, including varying \ce{CO2} concentrations (10–25 vol\%), total gas flow rates (50–100 mL min$^{-1}$), and adsorption temperatures (30–75 ºC). In parallel, molecular simulations of adsorption in representative activated carbon structures are combined with equilibrium mixture calculations and dynamic fixed-bed modeling. By directly comparing experimental breakthrough curves with model predictions, this study evaluates the ability of these models to describe dynamic adsorption behavior. In doing so, it provides mechanistic insight into the influence of pore structure, surface chemistry, and mass transfer on \ce{CO2} separation performance under practical operating conditions.

\section{Experimental methods}
\subsection{Activated carbons}
Two activated carbons with different origins and activation procedures were used in this work (Figure \ref{fig:S1}). A commercial activated carbon (WS-480), in extruded pellet form (4 mm diameter), was supplied by Calgon Carbon Corporation and produced from mineral coal through high-temperature steam activation. In addition, a biomass-derived activated carbon was produced in-house in a pilot-scale activation plant (Figure \ref{fig:S2}) from Miscanthus straw pellet biochar (MSP700), and denoted as MSP700-A900\ce{CO2}. The activation system consists of a vertical cylindrical ceramic fixed-bed reactor (120 cm length, 15 cm inner diameter) operating in downflow mode, equipped with a porous ceramic plate to support the carbon precursor (pellets). The reactor is heated by a three-zone electric furnace with independent temperature control, and the sample temperature is monitored using a K-type thermocouple inserted into the carbon bed. For the activation experiment, approximately 1 kg of biochar was loaded into the reactor. Prior to activation, the system was purged with \ce{N2} (5 L min$^{-1}$) for 60 min. The reactor was then heated at 4 ºC min$^{-1}$ under \ce{N2} flow up to 900 ºC. Once the target temperature was reached and stabilized, the gas was switched to pure \ce{CO2} (5 L min$^{-1}$) and maintained for 90 min. After activation, the sample was cooled down to room temperature under \ce{N2} flow (5 L min$^{-1}$). At the reactor outlet, a gas treatment system was installed, consisting of two jacketed stainless-steel condensers connected in series (cooled by a chiller at 0 ºC), along with a stainless-steel filter to retain volatile matter released from the biochar pellets before venting the outlet gases. In addition, the reactor is equipped with independent cooling circuits for the reactor heads (ambient temperature) and for vapor condensation (low-temperature circuit). 

The degree of activation, expressed as burn-off (wt.\%), was calculated according to Eq. \ref{eq:burn-off}: 

\begin{equation}\label{eq:burn-off}
    Burn-off\ (wt. \%)=\frac{W_{0,daf}-W_{f,daf}}{W_{0,daf}}\cdot100
\end{equation}

where $W_{0,daf}$ and $W_{f,daf}$ correspond to the dry ash-free mass of the biochar before and after activation, respectively.

\subsection{Physicochemical characterization of activated carbons}
Proximate analysis was performed according to European standards. Moisture content was determined by drying the sample at 105 ºC for 24 h (UNE-EN ISO 18134-1:2016). Volatile matter content was measured by thermogravimetric analysis (TGA) by heating the sample up to 900 ºC at 10 ºC min$^{-1}$ followed by a 20 min isothermal step under argon flow (UNE-EN ISO 18123:2016). Ash content was determined by combustion in a muffle furnace at 550 ºC for 120 min (UNE-EN ISO 18122:2016). The fixed carbon content (wt.\%) was calculated by difference: $Fixed\ Carbon\ (F.C.)=100-Moisture-Volatile\ matter-Ash$

Elemental composition was determined using a Thermo Scientific FLASH 2000 CHNS/O analyzer. Carbon, hydrogen, nitrogen, and sulfur contents were directly measured, while oxygen content (dry basis, wt.\%) was calculated by difference as follows: $O=100-C-H-N-S-Ash$

Ash samples were obtained by calcination of the carbon materials in a muffle furnace. The elemental composition of the ashes was determined by inductively coupled plasma optical emission spectroscopy (ICP-OES) using a PerkinElmer Optima 3300 DV instrument. Prior to analysis, the ash samples were digested in a mixture of \ce{HF} and \ce{HNO3} using an Anton Paar MW3000 microwave digestion system.

Textural properties were determined from \ce{N2} and \ce{CO2} adsorption-desorption isotherms measured at -196.15 and 0 ºC, respectively, using a Micromeritics 3Flex analyzer equipped with a high-vacuum system and 0.1 Torr pressure transducers. Samples were degassed at 140 ºC for 16 h prior to analysis. The specific surface area was calculated using the Brunauer-Emmett-Teller (BET) method, while the total pore volume was determined at a relative pressure of 0.97. Micropore volume and surface area ($V_{MIC}$ and $S_{MIC}$), as well as external surface area ($S_{EXT}$) were estimated using the t-plot method from \ce{N2} adsorption-desorption isotherm. The pore size distribution (PSD) was obtained by combining \ce{CO2} and \ce{N2} isotherms using SAIEUS software and applying a Two-Dimensional Non-Local Density Functional Theory (2D-NLDFT) model assuming slit-shaped pores.\cite{ishii_temperature_2016} Additional \ce{CO2} isotherms were measured at 15 and 25 ºC.

Fourier transform infrared (FTIR) spectra were recorded using a Thermo Fisher Scientific Nicolet 6700 spectrometer equipped with a ceramic source, KBr beam splitter, and DTGS detector. Spectra were collected in the range 4500-550 cm$^{-1}$ with a resolution of 4 cm$^{-1}$ and 32 scans. The spectra were processed using baseline and ATR corrections. Samples were ground into powder prior to analysis.

Temperature-programmed desorption (TPD) experiments were conducted using a Micromeritics AutoChem II 2910 instrument equipped with a thermal conductivity detector (TCD) and coupled to a Pfeiffer Omnistar GSD350 mass spectrometer. Approximately 100 mg of sample were loaded into a quartz U-shaped reactor and pretreated under helium flow at 140 ºC to remove physisorbed species. The temperature was then increased to 900 ºC at 10 ºC min$^{-1}$ under helium flow. The evolved gases (m/z = 18, 28, and 44, corresponding to \ce{H2O}, \ce{CO}, and \ce{CO2}) were continuously monitored.\cite{jagiello_enhanced_2019}

\subsection{Lab-scale fixed-bed \ce{CO2} adsorption tests}
The \ce{CO2} adsorption breakthrough experiments were carried out in a lab-scale fixed-bed setup, the schematic diagram of which is shown in Figure \ref{fig:S3}. The system consists of a vertical glass column (50 mm length, 16 mm inner diameter) equipped with a porous plate to support the adsorbent bed and a K-type thermocouple inserted into the bed to monitor the adsorption temperature. The adsorption tests were performed using 10 g of activated carbon pellets, previously dried at 110 ºC overnight to ensure complete moisture removal and degassing. The \ce{CO2}/\ce{N2} gas mixtures were fed in a downward flow configuration through the fixed bed. The influence of operating conditions was evaluated by varying the temperature (30, 50, and 75 ºC), \ce{CO2} concentration (10-25 vol\%), and total gas flow rate (50-100 mL min$^{-1}$).

Prior to each experiment, the reactor was purged with \ce{N2} at 100 mL min$^{-1}$ for 1 h to ensure an inert atmosphere. In addition, blank experiments were conducted under identical conditions to confirm negligible \ce{CO2} retention in the experimental setup. Subsequently, the gas mixture was directed to a bypass line connected to the gas chromatograph (GC) until a stable inlet composition was achieved. The flow was then switched to the adsorption column, and the outlet gas composition was continuously monitored by GC until saturation of the adsorbent bed was reached.

The amount of \ce{CO2} adsorbed was determined from the breakthrough curves by integrating the difference between inlet and outlet concentrations over time. The adsorption capacity, \textit{q} (mol g$^{-1}$), was calculated as (Eq. \ref{eq:ads_capacity}):

\begin{equation}\label{eq:ads_capacity}
    q=\frac{F}{m}\int_{0}^{t_s}(C_{in}-C_{out}(\text{t}))dt
\end{equation}

where $F$ is the total molar flow rate (mol s$^{-1}$), $m$ is the mass of adsorbent (g), $C_{in}$ and $C_{out}$(t) are the inlet and outlet \ce{CO2} concentrations, respectively, and $t_s$ (s) is the saturation time. The integral was evaluated numerically using the trapezoidal rule (Eq. \ref{eq:trapezoid}):

\begin{equation}\label{eq:trapezoid}
    q=\frac{F}{m}\sum_{i=1}^{n-1}\frac{[(C_{in}-C_{out,i})+(C_{in}-C_{out,i+1})]}{2}\cdot(t_{i+1}-t_{i})
\end{equation}

where $C_{out,i}$ is the outlet \ce{CO2} concentration at time $t_i$, and $n$ is the total number of experimental data points.

\subsection{Simulation details}
To be able to simulate adsorption and separation properties, we need periodic atomic models of the activated carbons to be used in the simulations. Experimentally, it is not possible to extract atomic positions from the structures, as activated carbons are amorphous materials. However, there have been modeling efforts towards generating realistic atomic representations of activated carbons, for instance, by performing hybrid reverse Monte-Carlo simulations or quenched molecular dynamics simulations.\cite{palmer_modeling_2010, bhatia_characterizing_2017} In 2020, Thyagarajan and Sholl reported a database of these computationally constructed activated carbon models, which we used for screening models for this work.\cite{thyagarajan_database_2020} To computationally investigate adsorption for WS-480, we used an atomistically realistic model of the porous saccharose coke CS1000a, reconstructed by Jain et al.\cite{jain_molecular_2006} WS-480 and CS1000a have been reported to have similar pore volumes and porosities.\cite{peng_adsorption_2017, wu_characterizations_2015} The reconstructed model of CS1000a has also been used to simulate adsorption isotherms of methanol, which showed good agreement with experimental isotherms of WS-480.\cite{madero-castro_adsorption_2022} For a description of the MSP700-A900\ce{CO2} structure, we tested several amorphous carbon structures reported in the database of Thyagarajan and Sholl.\cite{thyagarajan_database_2020} As MSP700-A900\ce{CO2} and WS-480 have similar pore size distributions in the ultramicropore regime, we searched for a model with a pore size distribution similar to that of CS1000a. After additionally comparing simulated isotherms with experimental data, we chose to use a model of an activated carbon fiber, reconstructed by Nguyen et al., which is denoted in literature as Bhatia 001.\cite{nguyen_new_2008}

Unlike WS-480 and MSP700-A900\ce{CO2}, both the chosen CS1000a and Bhatia models did not contain any oxygen containing functional groups. These groups can greatly influence \ce{CO2} adsorption, so we manually added functional groups to the CS1000a and Bhatia models. We randomly selected positions on the surface and placed either a carboxyl or hydroxyl group. The functional groups were relaxed after placement while keeping the framework atoms fixed, to ensure physically realistic structures. The relaxation was performed using the M3GNet-UP-2022 machine learned potential employed in the Amsterdam Modeling Suite.\cite{chen_universal_2022, baerends_amsterdam_2025} We added different amounts of functional groups to test how to most accurately reproduce the adsorption and separation properties of WS-480 and MSP700-A900\ce{CO2}. Because the CS1000a and Bhatia models have very different unit cell sizes, pore volumes, and surface areas, adding the same absolute number of groups to both models will have vastly different outcomes. We therefore tested adding the same amount of groups for both models either expressed in molar fraction (MF) or in number of groups per unit of surface area (SA). For both metrics, we subsequently selected the amount of groups that gave the best match with experimental isotherms. Detailed information on the tested amounts of added functional groups can be found in Section \ref{sec:AC_models_info}. The absolute amount of added functional groups for both models are listed in Table \ref{tab:no_func_groups}.

To calculate adsorption isotherms of \ce{CO2} and \ce{N2}, we performed grand canonical Monte-Carlo (GCMC) simulations using the RASPA3 software.\cite{ran_raspa3_2024} 10000 initialization and 50000 production cycles were performed for each pressure point. Both the AC frameworks and \ce{CO2} and \ce{N2} were treated as rigid, considering only van der Waals and electrostatic interactions. Shifted Lennard-Jones potentials without tail-corrections with a cutoff of 12 {\AA} were used to describe the van der Waals interactions. The Lennard-Jones parameters and charges for the AC models were taken from Jain et al. for the main framework atoms and from Jorge et al. for the functional group atoms.\cite{jain_molecular_2006, jorge_simulation_2002} For \ce{CO2} and \ce{N2}, the parameters were taken from García-Sanchez et al. and Martín-Calvo et al., respectively.\cite{garcia-sanchez_transferable_2009, martin-calvo_effect_2011} All parameters are listed in Table \ref{tab:ff_parameters}. Lorentz-Berthelot mixing rules were applied to compute cross-interactions.\cite{lorentz_ueber_1881, berthelot_sur_1898} The Ewald summation method was used to calculate the electrostatic interactions.\cite{darden_particle_1993} The heats of adsorption during the simulations were computed using a fluctuation formula.\cite{vlugt_computing_2008} We also computed the heats of adsorption from the experimental isotherms using the virial fit method.\cite{czepirski_virial-type_1989, nuhnen_practical_2020, lucassen_mixture_2026}

To model the \ce{CO2}/\ce{N2} separation with the activated carbon frameworks, we performed numerical simulations with the RUPTURA software.\cite{sharma_ruptura_2023} First, calculated adsorption isotherms were fitted to multi-site Sips equations (Equation \ref{eq:multi-sips}):\cite{sips_structure_1948}

\begin{equation}\label{eq:multi-sips}
    q(p)=\sum_{i} q_{i}^{\mathrm{sat}} \frac{{(b_{i}p)}^{1/\nu_i}}{1+(b_{i}p)^{1/\nu_i}}
\end{equation}

Here, \(q(p)\) is the absolute loading of the adsorbed phase, \(q_i^{sat}\) the saturation loading, \(p\) the pressure and \(b_i\) and \(\nu_i\) the affinity and heterogeneity factor, respectively. The resulting fitting parameters are listed in Tables \ref{tab:isotherm_fit_parameters_CO2} and \ref{tab:isotherm_fit_parameters_298_303_N2}. The fitted parameters were then used as input for the numerical simulations based on the Ideal Adsorption Solution Theory (IAST). We predicted multicomponent adsorption in equilibrium and calculated selectivities from the predicted adsorbed molar fractions. Dynamic separation was also predicted by performing fixed-bed breakthrough simulations using the RUPTURA software.\cite{sharma_ruptura_2023} Input parameters can be found in Section \ref{sec:breakthrough_info} and the column parameters used are listed in Table \ref{tab:column_parameters}.

\section{Results and Discussion}
\subsection{Characterization of activated carbons}
\label{sec:charac_ac}
\begin{figure*}[t]
    \centering
    \includegraphics[width=0.9\textwidth]{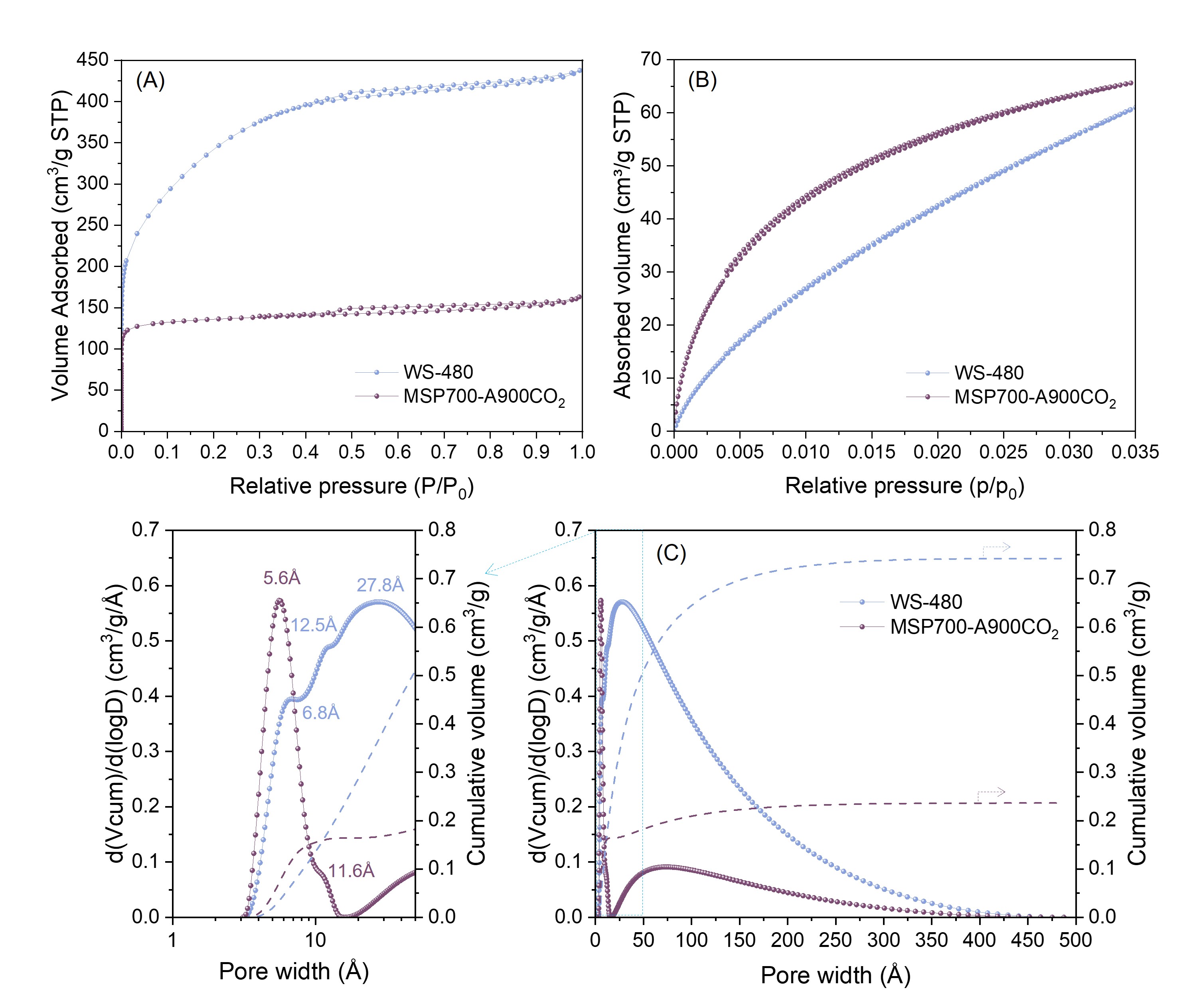}
    \caption{Adsorption-desorption isotherms of \ce{N2} at -196.15 ºC (A) and \ce{CO2} at 0 ºC (B) and 2D-NLDFT pore size distribution (C) of activated carbons (calculated from \ce{N2} and \ce{CO2} adsorption-desorption isotherms). }
    \label{fig:ads_des_PSD}
\end{figure*}

\begin{table*}[t]
    \caption{Textural properties determined from \ce{N2} isotherms of activated carbons}
    \centering
    \begin{tabular}{lccccc}
        \hline
        \multirow{2}{*}{Sample} 
            & $S_{BET}$ & $S_{MIC}$ & $S_{EXT}$ & $V_{MIC}$ & $V_{Total}$ \\
            & ($\mathrm{m^2/g_{db}}$) & ($\mathrm{m^2/g_{db}}$) 
            & ($\mathrm{m^2/g_{db}}$) & ($\mathrm{cm^3/g_{db}}$) 
            & ($\mathrm{cm^3/g_{db}}$) \\
        \hline
        WS-480 & 1188 & 981 & 207 & 0.49 & 0.67 \\
        \hline
        MSP700-A900\ce{CO2} & 529 & 437 & 92 & 0.258 & 0.185 \\
        \hline
    \end{tabular}
    \parbox{\linewidth}{\small $S_{BET}$: BET surface area; $S_{MIC}$ (micropore surface area), $S_{EXT}$ (external surface area) and $V_{MIC}$ (micropore volume) calculated using the t-plot method; $V_{Total}$: total pore volume at $P/P_0 \approx 0.97$. db: dry basis.}
    \label{tab:textural_properties}
\end{table*}

The physicochemical properties of the activated carbons were evaluated through a combination of textural, compositional, and surface characterization techniques in order to elucidate their influence on \ce{CO2} adsorption performance. From a textural standpoint, the \ce{N2} adsorption-desorption isotherms at -196.15 ºC (Figure \ref{fig:ads_des_PSD}A) and the derived parameters (Table \ref{tab:textural_properties}) indicate that the commercial activated carbon (WS-480) exhibits a significantly higher degree of porosity than the biomass-derived material (MSP700-A900\ce{CO2}). In particular, WS-480 shows higher values of specific surface area ($S_{BET}$), micropore surface area ($S_{MIC}$), micropore volume ($V_{MIC}$), and total pore volume ($V_{Total}$). This behavior is consistent with its higher \ce{N2} uptake over the entire relative pressure range, suggesting a more developed pore network with a substantial contribution from both micropores and wider pores. 

However, this trend contrasts with the \ce{CO2} adsorption isotherms at 0 ºC (Figure \ref{fig:ads_des_PSD}B), where the biomass-derived activated carbon exhibits higher \ce{CO2} uptake at low pressures. This apparent discrepancy highlights a well-known limitation of \ce{N2} adsorption at -196.15 ºC for characterizing ultramicroporosity, as diffusion limitations may hinder the access of \ce{N2} molecules to pores below $\sim$0.7 nm.\cite{silvestre-albero_physical_2012, thommes_physisorption_2015} 

\begin{table*}[t]
    \centering
    \caption{Proximate and ultimate analysis of activated carbons}
    \begin{tabular}{lccccccccc}
        \hline
        \multirow{3}{*}{Sample} 
            & \multirow{3}{*}{\begin{tabular}{c}Burn-off\\(wt.\%)\end{tabular}} 
            & \multicolumn{3}{c}{Proximate Analysis} 
            & \multicolumn{5}{c}{Ultimate Analysis} \\
            & 
            & \multicolumn{3}{c}{db (wt.\%)} 
            & \multicolumn{5}{c}{daf (wt.\%)} \\
        \cline{3-10}
            &
            & \begin{tabular}{c}Volatile\\Matter\end{tabular}
            & Ash 
            & \begin{tabular}{c}Fixed\\Carbon\end{tabular}
            & C & H & N & S & O \\
        \hline
        WS-480 
            & -- 
            & 3.3 & 12.1 & 84.6 
            & 92.7 & 0.5 & 0.3 & 0.3 & 6.2 \\
        MSP700-A900\ce{CO2} 
            & 26.0 
            & 0.0 & 25.0 & 75.0 
            & 94.9 & 0.6 & 0.6 & 0.0 & 3.9 \\
        \hline
    \end{tabular}
    \begin{flushleft}
    \small db: dry basis; daf: dry and ash free basis.
    \end{flushleft}
    \label{tab:prox_ult_analysis}
\end{table*}

\begin{figure*}[t]
    \centering
    \includegraphics[width=0.8\textwidth]{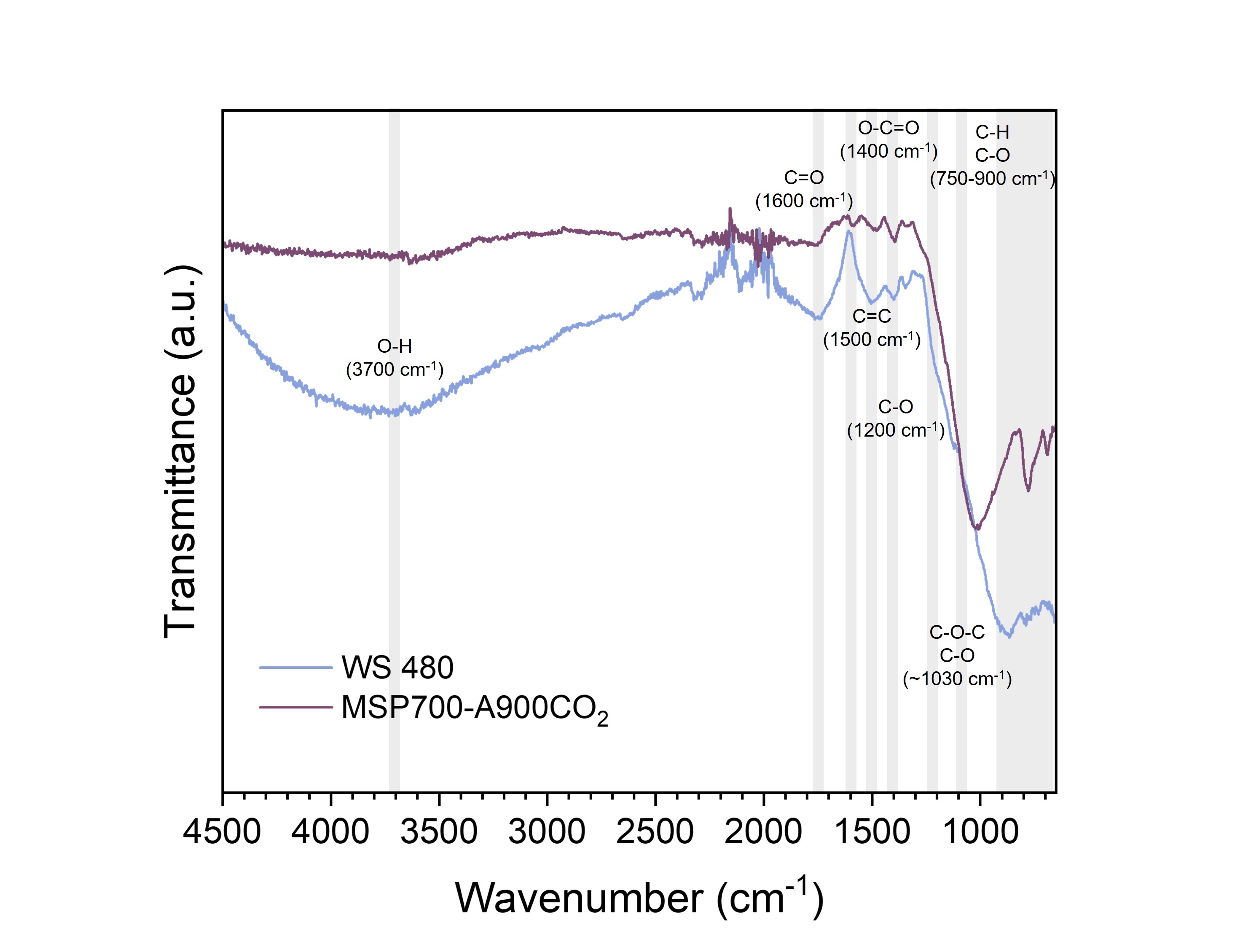}
    \caption{FTIR spectra of WS-480 and MSP700-A900\ce{CO2}.}
    \label{fig:FTIR}
\end{figure*}

This interpretation is further supported by the pore size distribution obtained from 2D-NLDFT analysis (Figure \ref{fig:ads_des_PSD}C), which reveals that MSP700-A900\ce{CO2} exhibits a more pronounced contribution in the ultramicropore region ($<$0.7 nm), whereas WS-480 displays a broader pore size distribution extending toward larger micropores. The importance of ultramicropores in enhancing \ce{CO2} adsorption has been extensively reported, particularly for pores in the range of 0.4-0.7 nm due to the overlapping adsorption potentials within confined spaces.\cite{sevilla_sustainable_2011, presser_effect_2011}

Beyond textural properties, the bulk composition of the carbons also reveals relevant differences between both materials. As shown in Table \ref{tab:prox_ult_analysis}, the biomass-derived activated carbon presents a higher carbon content and slightly higher nitrogen content, together with a lower oxygen content, than the commercial sample. In addition, proximate analysis indicates that MSP700-A900\ce{CO2} contains a significantly higher ash content than WS-480. These results suggest that both carbons differ not only in pore development but also in carbon matrix composition and inorganic fraction, which may influence their final adsorption behavior.

Although MSP700-A900\ce{CO2} exhibits a lower bulk oxygen content, FTIR and TPD-MS analyses indicate that the nature, distribution, and thermal stability of the surface oxygen functionalities differ markedly between both carbons. The FTIR spectra (Figure \ref{fig:FTIR}) display characteristic bands assigned to O–H stretching around 3700 cm$^{-1}$, mainly associated with isolated hydroxyl groups, aromatic C=C vibrations in the 1600-1500 cm$^{-1}$ region, C–O-containing functionalities such as phenols, ethers, and lactones between 1200 and 1030 cm$^{-1}$, and out-of-plane aromatic C–H vibrations in the 750-900 cm$^{-1}$ region, confirming the predominantly aromatic character of both carbons together with the presence of oxygenated surface species. 

\begin{figure}[t]
    \centering
    \includegraphics[width=0.5\textwidth]{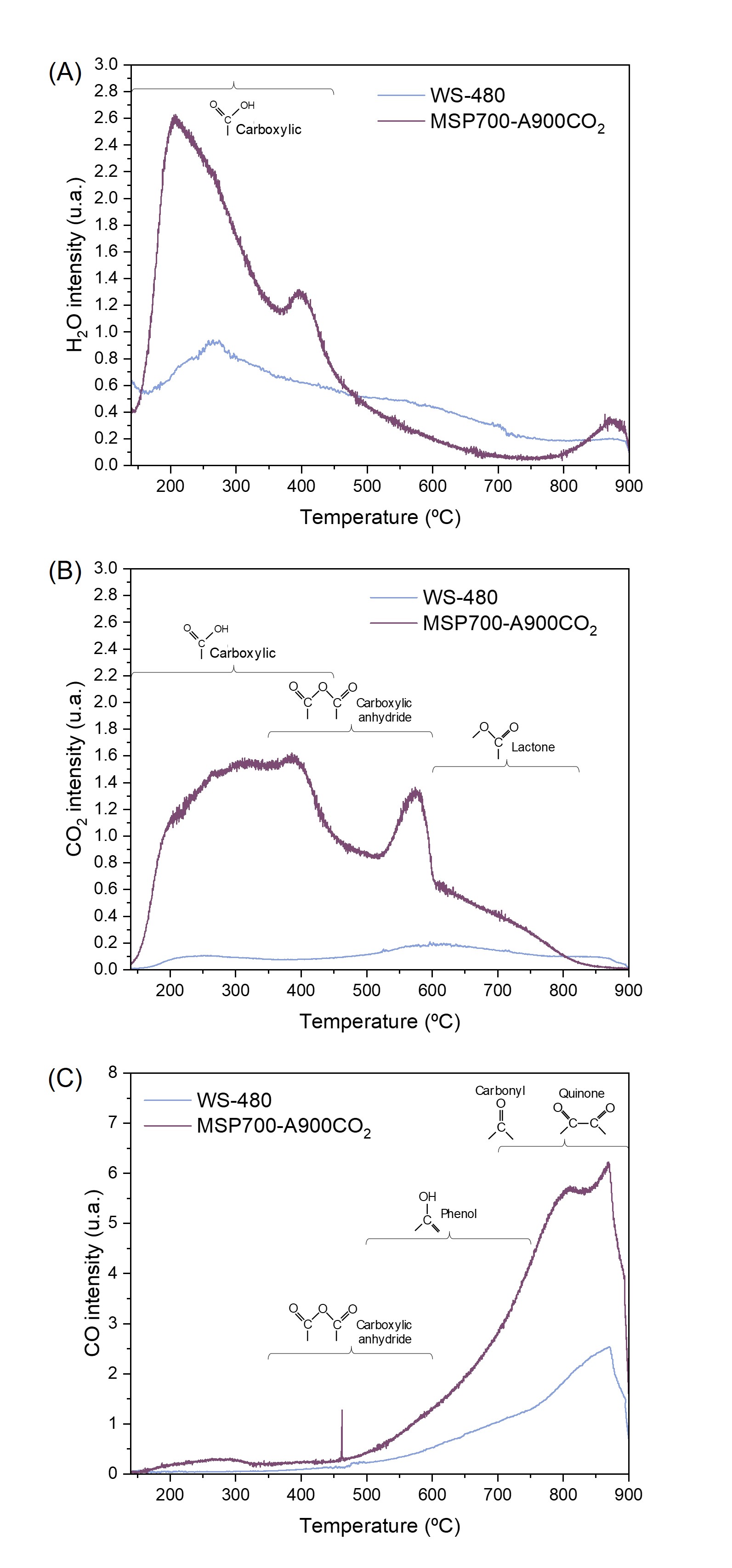}
    \caption{\ce{H2O} (A), \ce{CO2} (B), and \ce{CO} (C) TPD-MS spectra of activated carbons}
    \label{fig:TPD-MS}
\end{figure}

Despite the similar families of functionalities identified by FTIR (Figure \ref{fig:FTIR}), the TPD-MS profiles (Figure \ref{fig:TPD-MS}) reveal important differences in their abundance and thermal behavior. In particular, MSP700-A900\ce{CO2} shows \ce{H2O} evolution at approximately 200, 400, and 875 ºC, suggesting the contribution of weakly retained water and/or dehydration of labile oxygen-containing surface species, as well as more stable hydroxyl-containing structures. By contrast, WS-480 exhibits a narrower \ce{H2O} evolution peak centered at 260-275 ºC, indicative of a more limited distribution of water-releasing surface species. This apparent discrepancy is not necessarily contradictory, since the FTIR band near 3700 cm$^{-1}$ mainly reflects isolated O–H groups (Figure \ref{fig:FTIR}), whereas \ce{H2O} evolution during TPD-MS may also arise from physisorbed or strongly retained water and from dehydration reactions involving other oxygen-containing functionalities during heating.

The \ce{CO2} evolution profiles also differ significantly between both materials (Figure \ref{fig:TPD-MS}B). MSP700-A900\ce{CO2} exhibits several \ce{CO2} desorption peaks at approximately 300, 385, and 575 ºC, which are commonly attributed to the decomposition of carboxylic acids, anhydrides, and lactone-type groups. In contrast, WS-480 shows only two low-intensity \ce{CO2} evolution peaks at around 250 ºC and 615-625 ºC, indicating a simpler distribution and lower abundance of acidic oxygen-containing surface functionalities.\cite{figueiredo_modification_1999, boehm_surface_2002} The \ce{CO} evolution profiles reveal a broad signal extending from approximately 500 ºC to 900 ºC, with a maximum centered at 800-850 ºC (Figure \ref{fig:TPD-MS}C). This high-temperature region is typically associated with the decomposition of thermally stable oxygen-containing groups such as phenols, ethers, carbonyls, and quinones.\cite{figueiredo_modification_1999, boehm_surface_2002} In the case of MSP700-A900\ce{CO2}, this signal is markedly more intense than in WS-480, indicating a higher concentration of strongly bound oxygenated functionalities. Overall, the TPD-MS results indicate that the biomass-derived activated carbon exhibits a broader and more heterogeneous distribution of oxygen-containing surface groups over a wider range of thermal stabilities. Such differences in surface chemistry are relevant for \ce{CO2} adsorption, since the type and distribution of oxygen functionalities can affect the local polarity, acid-base character, and interaction energy of the carbon surface. Accordingly, the broader functional heterogeneity observed for MSP700-A900\ce{CO2} may generate a wider distribution of adsorption energies, thereby influencing \ce{CO2}-surface interactions under dynamic conditions.\cite{kundu_comprehensive_2024, petrovic_impact_2022, chen_complete_2024}

\begin{table*}[t]
    \caption{Mineral matter composition of commercial activated carbon WS-480 and biochar MSP700 ash}
    \centering
    \begin{tabular}{lcccccccccc}
        \hline
        Element, db & \multirow{2}{*}{Al} & \multirow{2}{*}{Ca}
        & \multirow{2}{*}{Fe} & \multirow{2}{*}{K} & \multirow{2}{*}{Mg}
        & \multirow{2}{*}{Na} & \multirow{2}{*}{P} & \multirow{2}{*}{Ti}
        & Others & \multirow{2}{*}{Ash} \\
        (\%wt) & & & & & & & & & (\%wt$<$0.1) \\
        \hline
        WS-480 & 3.4 & 2.5 & 3.8 & 0.3 & 0.7 & 0.9 & 0.0 & 0.2 & 0.3 & 12.1 \\
        \hline
        MSP700 & 0.31 & 4.30 & 0.66 & 10.25 & 1.39 & 0.57 & 1.42 & 0.0 & 0.20 & 19.1 \\
        \hline
    \end{tabular}
    \begin{flushleft}
    \small db: dry basis
    \end{flushleft}
    \label{tab:mineral_composition}
\end{table*}

The inorganic fraction may also have played an important role during the activation process. Although a high ash content is often associated with dilution effects or partial pore blockage, the ash composition of the MSP700 precursor (Table \ref{tab:mineral_composition}) reveals a high content of alkali and alkaline earth metals, especially K and Ca, which are known to catalyze gasification reactions and to promote the development of microporosity during \ce{CO2} activation.\cite{perander_catalytic_2015, romero_millan_catalytic_2019} Therefore, the higher mineral content of the biomass-derived precursor may have contributed to the formation of a narrower and more \ce{CO2}-relevant microporous structure, even if the final total porosity derived from \ce{N2} adsorption remained lower than that of WS-480.

\subsection{Pure and multi-component adsorption modeling}
\label{sec:pure_multi_modeling}
\begin{figure}[h!]
    \centering
    \includegraphics[]{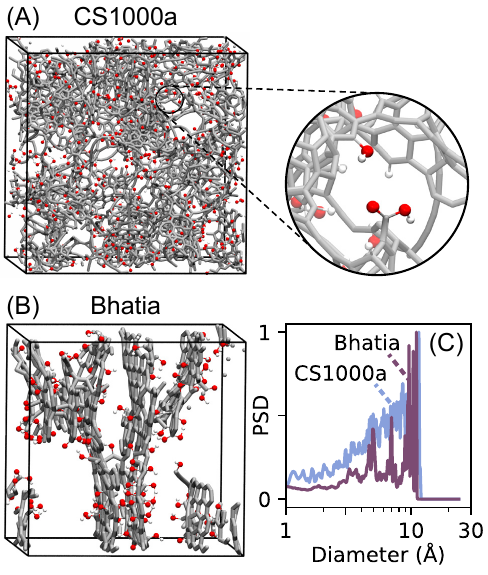}
    \caption{Snapshots of the unit cells of the CS1000a (A) and Bhatia (B) model. The zoomed-in part of the CS1000a snapshot shows an added carboxyl and hydroxyl group on the surface. Colors: C (grey), O (red), H (white). (C) Simulated pore size distributions of the CS1000a and Bhatia models without added functional groups.} 
    \label{fig:characterization_models}
\end{figure}

For the computational modeling, two existing activated carbon models were used as representations of WS-480 and MSP700-A900\ce{CO2}. The models will be denoted as CS1000a and Bhatia, respectively. Snapshots of the unit cells show substantial structural differences between the CS1000a and Bhatia models (Figures \ref{fig:characterization_models}A and B). The CS1000a model consists of clusters of small carbon sheets that are often twisted, creating highly irregular surfaces and pore geometries. In contrast, the Bhatia model is constructed from a slit-pore geometry, resulting in a more uniform surface. The pore size distributions further highlight these structural differences, with the Bhatia model exhibiting more distinct peaks, while the CS1000a model shows a more uniform pore size distribution, reflecting its highly disordered structure (Figure \ref{fig:characterization_models}C). Carboxyl and hydroxyl groups were added at random positions on the structures and were subsequently relaxed to ensure physically correct geometries, as shown in the zoomed-in snapshot in Figure \ref{fig:characterization_models}A. 

\begin{figure*}[t]
    \centering
    \includegraphics[]{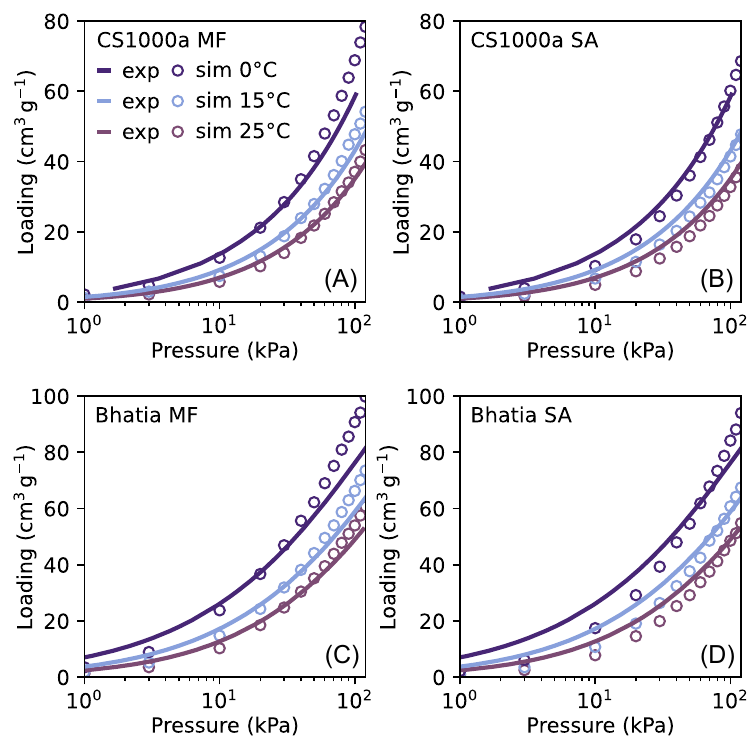}
    \caption{Experimental isotherms of \ce{CO2} in WS-480 (A, B) and MSP700-A900\ce{CO2} (C, D) compared to simulated models with 9 \% (MF) and 1.5-2 $\mathrm{\mu}$mol $\mathrm{m^{-2}}$ (SA) added functional groups.}
    \label{fig:exp_sim_isotherms}
\end{figure*}

Adsorption isotherms of \ce{CO2} in both WS-480 and MSP700-A900\ce{CO2} were measured at different temperatures. These isotherms are plotted in Figure \ref{fig:exp_sim_isotherms} together with simulated adsorption isotherms obtained using the CS1000a and Bhatia models, respectively, with added functional groups. To systematically evaluate the number of groups added to both models, we tested different molar fractions and groups per unit surface area. Ultimately, adding 9 \% functional groups and 1.5–2 $\mathrm{\mu}$mol $\mathrm{m^{-2}}$ resulted in the best agreement with the experimental isotherms. For future reference, these amounts will be denoted either by molar fraction (MF) or surface area (SA). The absolute amount of functional groups is higher in both MF models than in the SA models.

The \ce{CO2} loadings are higher for the MF models over the entire range shown in Figure \ref{fig:exp_sim_isotherms}, indicating stronger interactions while the systems are still far from saturation. This behavior can be explained by the greater number of functional groups in the MF models, which are expected to increase the interaction between \ce{CO2} and the frameworks. Between the MF and SA models, at lower pressures ($<$ 30 kPa), the MF models better reproduce the \ce{CO2} loadings of the experimental samples (Figures \ref{fig:exp_sim_isotherms}A and C). Around 100 kPa, however, the simulated isotherms of the MF models begin to overshoot the experimental isotherms, while the SA models show better agreement. 

Functional groups are expected to increase the interaction between \ce{CO2} and the frameworks, while simultaneously decreasing the available pore volume. This behavior translates into higher initial loadings but lower saturation loadings of \ce{CO2} when more functional groups are present. The \ce{CO2} loadings are higher for the MF models over the entire range shown in Figure \ref{fig:exp_sim_isotherms}, due to the increased interactions with \ce{CO2} and the systems still being far from saturation. Between the MF and SA models, at lower pressures ($<$ 30 kPa), the MF models better represent the \ce{CO2} loadings of the experimental samples (Figures \ref{fig:exp_sim_isotherms}A and C). Around 100 kPa, however, the simulated isotherms of the MF models begin to overshoot the experimental isotherms, while the SA models show better agreement. 

Still, for both CS1000a models, the agreement with the experimental isotherms of WS-480 is very good, and the isotherms shapes closely match each other (Figures \ref{fig:exp_sim_isotherms}A and B). Additionally, the calculated heats of adsorption from the experimental isotherm closely agree with those obtained from the simulations (Figure S4A). This is consistent with previously reported behavior, where simulations using the CS1000a model were able to reproduce experimental isotherms of WS-480 for methanol.\cite{madero-castro_adsorption_2022} 

In contrast, the shapes of the isotherms obtained with the Bhatia models differ more from those of MSP700-A900\ce{CO2} (Figures \ref{fig:exp_sim_isotherms}C and D). Both Bhatia MF and SA models begin to overestimate the experimental \ce{CO2} loadings at slightly lower pressures, while underestimating the loading at low pressures to a greater extent. At low pressures, the simulated heats of adsorption also strongly underestimate the experimental heats of adsorption of MSP700-A900\ce{CO2} (Figure S4B). The experimental heat of adsorption then decreases and converges toward the simulated values around 2 mmol g$^{-1}$. 

The simulated heat-of-adsorption profiles of the Bhatia models are more linear, suggesting that preferential adsorption sites are absent in the Bhatia frameworks, whereas the structure of MSP700-A900\ce{CO2} most likely contains such preferential sites. This indicates that the pore structure of MSP700-A900\ce{CO2} deviates more from the model used in the simulations than WS-480 does from the CS1000a model. Additionally, in Section \ref{sec:charac_ac}, we highlighted that MSP700-A900\ce{CO2} exhibits broader functional heterogeneity than WS-480, while in this study only carboxyl and hydroxyl groups were added to the models. 

Still, the absolute difference in loading between the experiments and the Bhatia models is at most 18 cm$^{3}$ g$^{-1}$ at 120 kPa and 0 ºC. At room temperature, the differences are even smaller, indicating that the Bhatia models are able to reproduce the adsorption of \ce{CO2} in MSP700-A900\ce{CO2} well. These results highlight that our strategy of screening existing activated carbon models from the literature and adding functional groups to their surfaces is effective for modeling the adsorption behavior of an activated carbon with an unknown structure.

\begin{figure*}[t]
    \centering
    \includegraphics[]{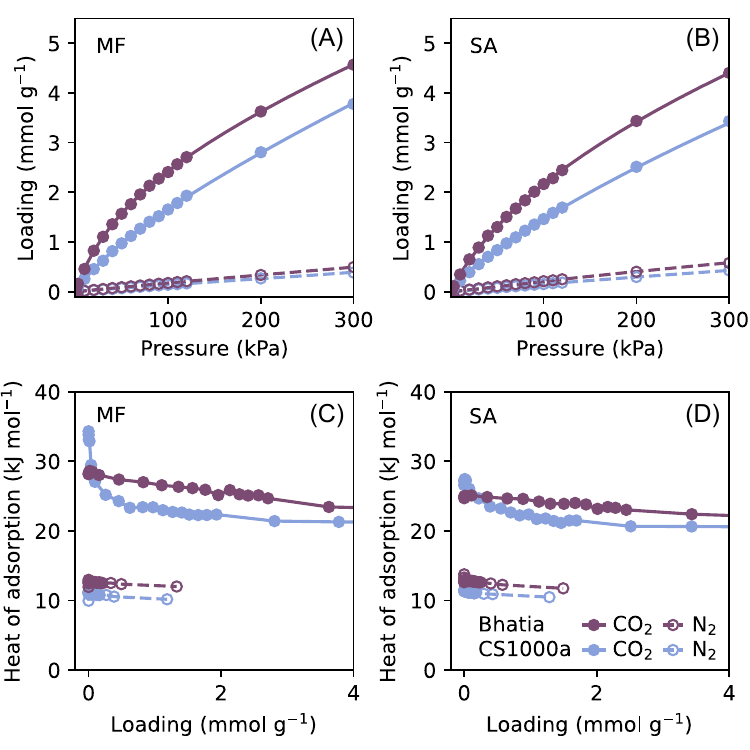}
    \caption{Simulated adsorption isotherms (A, B) and heats of adsorption (C, D) of \ce{CO2} and \ce{N2} at 25 ºC in the MF (A, C) and SA (B, D) CS1000a and Bhatia models. In the isotherms, the curves are fitted multi-site Sips equations, of which the parameters are listed in Tables S3 and S4.} 
    \label{fig:N2_heats_isotherms}
\end{figure*}

Now that the computational activated carbon models and the force field have been validated, we extend the pressure range to 300 kPa and compute the \ce{N2} adsorption isotherms in the four models. The pure component isotherms and heats of adsorption at 25 ºC are shown in Figure \ref{fig:N2_heats_isotherms}.

\begin{figure*}[t]
    \centering
    \includegraphics[]{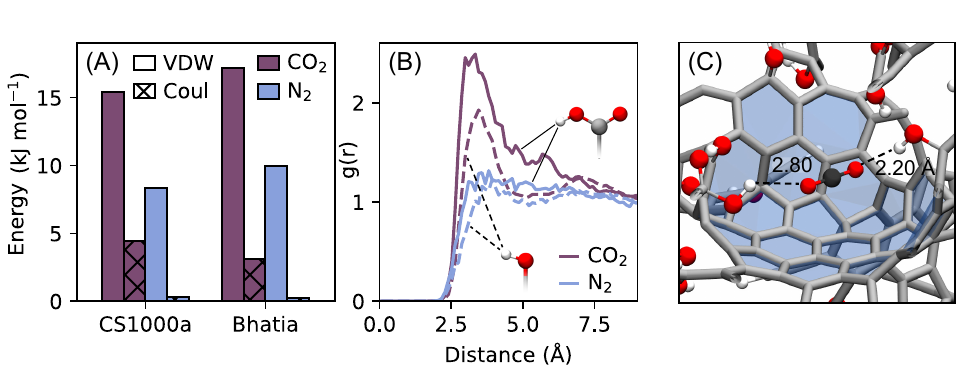}
    \caption{(A) Average van der Waals (VDW) and Coulombic (Coul) energy per adsorbate molecule with the framework of the SA models at 100 kPa. (B) Radial distribution functions of the hydrogen atoms of the carboxyl and hydroxyl groups with the center of mass of \ce{CO2} and \ce{N2} in CS1000a SA at 100 kPa. (C) Snapshot of the GCMC simulations of an adsorbed \ce{CO2} molecule in the CS1000a MF model at 0.1 kPa. The curved carbon sheet around \ce{CO2} is colored blue, with attached two functional groups close to \ce{CO2}.} 
    \label{fig:adsorption_analysis}
\end{figure*}

In all models, both the simulated adsorbed loadings and the heats of adsorption of \ce{N2} are significantly lower than those of \ce{CO2} (Figure \ref{fig:N2_heats_isotherms}). At 100 kPa, the loading of \ce{N2} does not exceed 0.18 mmol g$^{-1}$, and at 300 kPa it increases only to 0.5 mmol g$^{-1}$. The large difference between the adsorption properties of \ce{CO2} and \ce{N2} is also reflected in their heats of adsorption, with values for \ce{CO2} being at least 15 kJ mol$^{-1}$ higher. This large energetic difference is mainly caused by the higher polarizability of \ce{CO2}, which gives rise to stronger van der Waals interactions between \ce{CO2} and the framework (Figure \ref{fig:adsorption_analysis}A).

Additionally, the higher quadrupole moment of \ce{CO2} increases the Coulombic interactions with the functional groups compared to \ce{N2} (Figure \ref{fig:adsorption_analysis}A). This stabilization is visible in the radial distribution functions of the hydrogen atoms of both types of functional groups in CS1000a. \ce{CO2} exhibits a peak around 3 {\AA}, whereas this peak is absent for \ce{N2} (Figure \ref{fig:adsorption_analysis}B). The reason for this favorable interaction with \ce{CO2} lies in the Lewis acidic sites of the carboxyl and hydroxyl groups, which interact with the oxygen atoms of \ce{CO2}. The heats of adsorption of \ce{CO2} therefore increase with increased surface functionalization, as evidenced by the higher heats of adsorption for the MF models (Figures \ref{fig:N2_heats_isotherms}C and D). 

Finally, comparing the Bhatia with the CS1000a models, the Bhatia models generally exhibit higher adsorbed loadings and heats of adsorption (Figures \ref{fig:N2_heats_isotherms}C and D). Only at very low \ce{CO2} loadings are the heats of adsorption higher for the CS1000a models. This behavior arises because the Bhatia models show only a slight, gradual decline in the heat of adsorption, whereas both CS1000a models exhibit a steeper decrease at low \ce{CO2} loadings. This effect is most pronounced for the MF variant of CS1000a, in which the heat of adsorption starts at 34 kJ mol$^{-1}$ and decreases to around 25 kJ mol$^{-1}$ by approximately 0.5 mmol g$^{-1}$. A much smaller decline is observed for the SA model. 

To investigate the reason for the sharp decline in the heat of adsorption for the MF model of CS1000a, we visually inspected snapshots from the GCMC simulations at 0.1 kPa. One such snapshot is shown in Figure \ref{fig:adsorption_analysis}C. In the MF model, a few sites exist where multiple functional groups are located close to one another within a confined space created by a highly curved carbon sheet, thereby providing preferential adsorption sites for \ce{CO2}. Once these sites are occupied, the heat of adsorption drops below that of the Bhatia models. The more linear heats-of-adsorption profiles observed for \ce{CO2} in the Bhatia models, indicative of energetically more equivalent adsorption sites, arise from the more uniform surface of these models (Figure \ref{fig:characterization_models}B). Moreover, aside from the first few molecules entering the framework, the Bhatia models exhibit stronger interactions with \ce{CO2}. This trend is consistent with the experimental isotherms of WS-480 and MSP700-A900\ce{CO2}, which show higher \ce{CO2} loadings for the MSP700-A900\ce{CO2} framework (Figure \ref{fig:exp_sim_isotherms}). 

\begin{figure*}[t]
    \centering
    \includegraphics[]{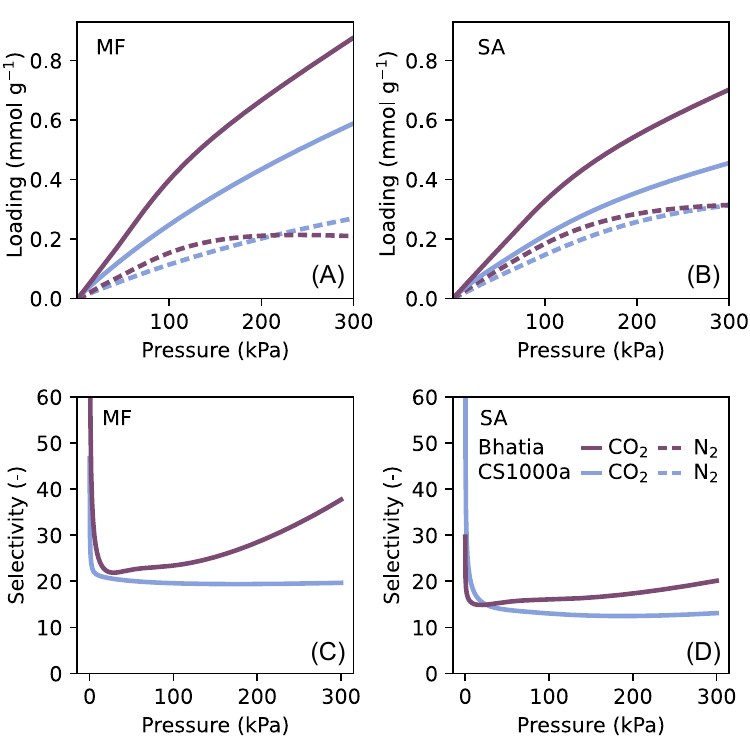}
    \caption{Simulation-based predictions of the separation of a 10:90 \ce{CO2}/\ce{N2} mixture at 25 ºC in the MF (A, C) and SA (B, D) models of CS1000a and Bhatia. (A, B) Multi-component adsorption isotherms (C, D) Selectivities.} 
    \label{fig:IAST}
\end{figure*}

As the pure component isotherms predict preferential adsorption of \ce{CO2} over \ce{N2}, we investigate the separation performance, first under equilibrium conditions. We used the fitted Sips equations as input for calculations based on IAST and determined selectivities from the predicted adsorbed molar fractions. The results for the separation of a 10:90 \ce{CO2}/\ce{N2} mixture at 25 ºC in the four activated carbon models are shown in Figure \ref{fig:IAST}. 

The highest selectivities are predicted for the Bhatia MF model, reaching approximately 40 at 300 kPa. However, at more representative conditions (\ce{CO2}/\ce{N2} 10:90, 25 ºC), the selectivity is approximately 23. Reported IAST selectivities for other inorganic carbon adsorbents under the same conditions cover a wide range, from, 5 to 50, depending on the modifications made to the frameworks, with typical values between 10 and 30.\cite{zhao_microporous_2015} The predicted selectivities are therefore in line with those of comparable microporous carbon materials. 

Due to the low partial pressure of \ce{CO2} in flue gas, the absolute \ce{CO2} loading in the activated carbon models is expected to be below 1 mmol g$^{-1}$ (Figures \ref{fig:IAST}A and B). However, this causes the relative effect of the functional groups to be much more pronounced. At 100 kPa, the difference in \ce{CO2} loading between the MF and SA models is approximately 20\% for both CS1000a and Bhatia. As a result, combined with a predicted decrease in \ce{N2} loading, the selectivities are lower for the SA models (Figures \ref{fig:IAST}C and D). A higher concentration of carboxyl and hydroxyl functional groups is therefore predicted to be beneficial for increasing the selectivity in activated carbon structures. These results highlight that these computational methods enable efficient evaluation of the influence of different structural properties of activated carbons on separation performance.

\subsection{Fixed-bed breakthrough experiments and simulations}
\begin{figure*}[t]
    \centering
    \includegraphics[]{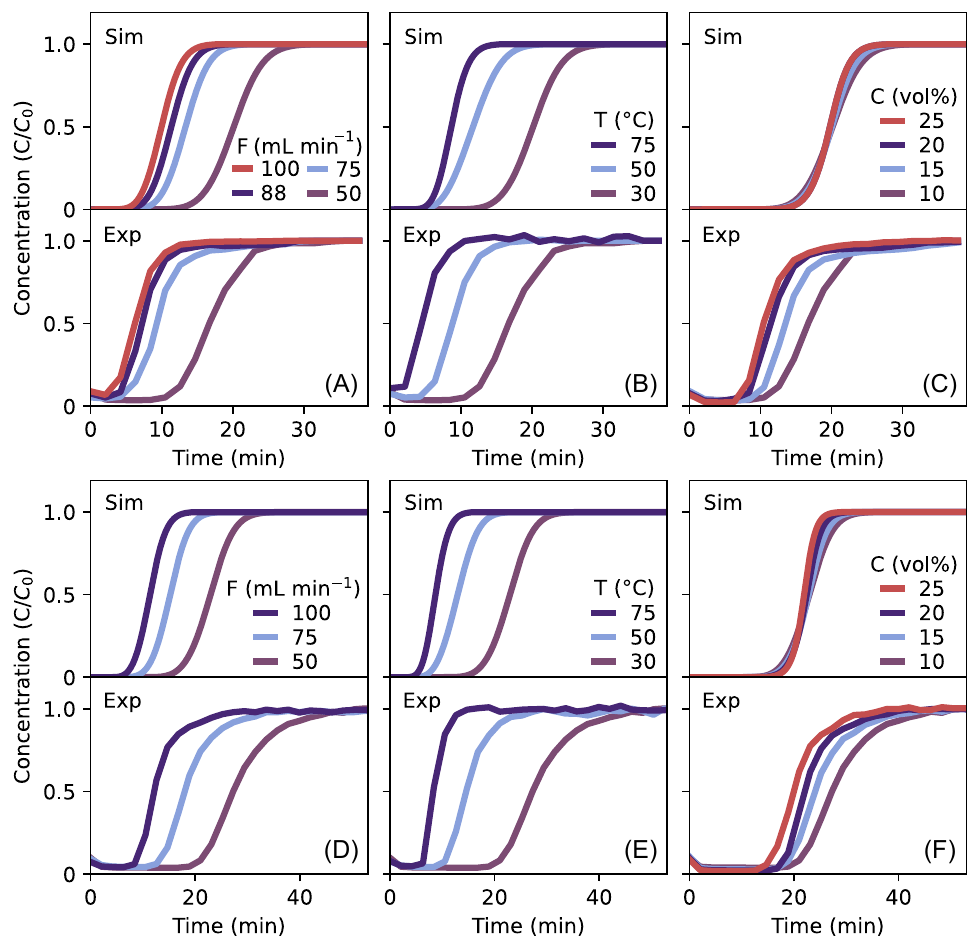}
    \caption{Simulated and experimental breakthrough curves of the SA models of CS1000a/WS-480 (A,B,C) and Bhatia/MSP700-A900\ce{CO2} (D,E,F) at different flow rates (A,D), temperatures (B,E), and \ce{CO2}-concentrations (C,F). The simulated curves of \ce{N2} are omitted for clarity.} 
    \label{fig:breakthroughs}
\end{figure*}

For the separation of a 10:90 \ce{CO2}/\ce{N2} mixture, the Bhatia models predict significantly higher \ce{CO2} loadings than the CS1000a models, along with a modest improvement in \ce{CO2}/\ce{N2} selectivity. This trend aligns with the obtained experimental isotherms: the Bhatia model captures the stronger \ce{CO2} adsorption observed for MSP700-A900\ce{CO2}, whereas the CS1000a model reflects the comparatively weaker adsorption behavior of WS-480. The next step is to evaluate the \ce{CO2} adsorption performance of the activated carbons under dynamic conditions in a fixed-bed system. The breakthrough curves from the fixed-bed experiments are shown together with the simulated curves of \ce{CO2} in Figure \ref{fig:breakthroughs}. 

Simulated breakthrough curves of \ce{CO2} in the SA models of CS1000a and Bhatia show close agreement with those of WS-480 and MSP700-A900\ce{CO2} at different flow rates, temperatures, and concentrations (Figure \ref{fig:breakthroughs}). The trends are mostly captured qualitatively, and the breakthrough times also agree well with the experiments. The main discrepancy between the simulations and experiments occurs for the different \ce{CO2} concentrations (Figures \ref{fig:breakthroughs}C and F) and arises from an underestimation of the width of the mass-transfer zone in the simulations. The width of the mass-transfer zone is influenced by many factors, so the exact cause of this difference is difficult to determine. However, the simulation model employs several approximations, such as assuming isothermal conditions, which have been reported to result in an underestimation of the mass-transfer zone width.\cite{sharma_ruptura_2023, gooijer_tamof-1_2025} In particular, for \ce{CO2} adsorption, which is an exothermic process, thermal gradients in the column can cause the adsorption front to spread. Nevertheless, the numerical model is able to closely capture the dynamic adsorption behavior in a fixed-bed system for both activated carbons.

\begin{figure*}[t]
    \centering
    \includegraphics[]{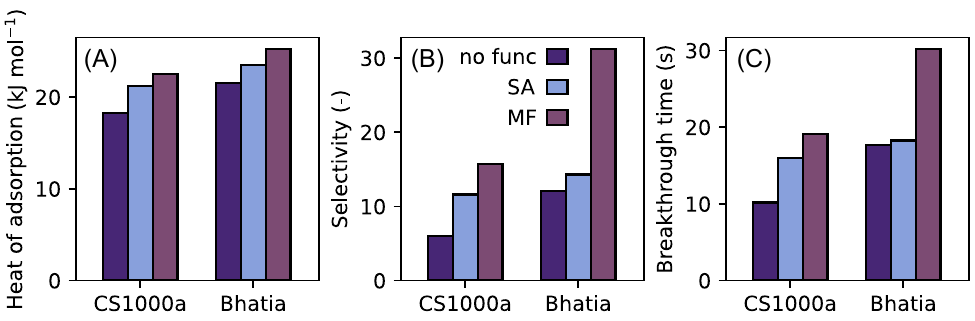}
    \caption{(A) Heats of adsorption at 30 ºC and 100 kPa. Predicted selectivities (B) and breakthrough times (C) of \ce{CO2} from a 10:90 \ce{CO2}/\ce{N2} mixture at 30 ºC and 100 kPa.} 
    \label{fig:func_groups_analysis}
\end{figure*}

In Section \ref{sec:pure_multi_modeling}, we discussed the effect of functionalization of the activated carbon surface on the adsorption properties of \ce{CO2} and \ce{N2}. Now that the dynamic separation of \ce{CO2} and \ce{N2} has also been simulated, we can analyze the influence of functional groups in more detail. The heats of adsorption, selectivity, and predicted breakthrough times of \ce{CO2} for all simulated models, including those without added functional groups, are plotted in Figure \ref{fig:func_groups_analysis}. 

The favorable interaction between \ce{CO2} and the functional groups is clearly reflected in increased heats of adsorption, selectivities, and breakthrough times for \ce{CO2} with increasing functionalization (Figure \ref{fig:func_groups_analysis}). There are, however, differences in the extent to which these properties are affected. For both the CS1000a and Bhatia models, the heat of adsorption increases incrementally with by only a few kJ mol$^{-1}$. This relatively small difference in heat of adsorption nevertheless causes large relative changes in selectivity, with the most striking example being the predicted selectivity of the SA and MF variants of the Bhatia model, which increases with a factor of 2.2. The trends of the selectivity are similar to those observed for the retention times in the breakthrough simulations, with the largest difference again occurring between the SA and MF variants of the Bhatia model. These differences in breakthrough times, however, are not solely due to differences in selectivity. With increased functionalization, the particle density and column void fraction also change, more prominently for the Bhatia models, which can lead to different variations in breakthrough times. Previous work has shown that column conditions strongly influence the sensitivity of breakthrough times to fluctuations.\cite{gooijer_tamof-1_2025}

The \ce{CO2} adsorption capacities obtained from breakthrough experiments under different operating conditions are summarized in Table \ref{tab:ads_capacities_conditions}. As expected, the adsorption capacity increased with increasing \ce{CO2} concentration for both activated carbons, reflecting the higher driving force for mass transfer at higher partial pressures. In contrast, an increase in adsorption temperature led to a noticeable decrease in \ce{CO2} uptake, which is consistent with the exothermic nature of the adsorption process. Regarding the effect of gas flow rate, a slight decrease in adsorption capacity was observed at higher flow rates, which can be attributed to reduced gas-solid contact time and incomplete utilization of the adsorbent bed. 

\begin{table*}[t]
    \centering
    \caption{\ce{CO2} adsorption capacities of activated carbons under different operating conditions}
    \begin{tabular}{lccc}
        \hline
        \multicolumn{2}{c}{\multirow{2}{*}{\textbf{Operating conditions}}} & \multicolumn{2}{c}{\textbf{Adsorption capacity (mmol g$\mathbf{^{-1}}$)*}}\\
        \cline{3-4}
        & & WS-480 & MSP700-A900\ce{CO2} \\
        \hline
        $\mathbf{Q_T}$ \textbf{(mL min$\mathbf{^{-1}}$)}  & 50 & 0.370 & 0.640 \\
        \ce{CO2} = 10 vol\% & 75 & 0.338 & 0.660 \\
        $T$ = 30 ºC & 88 & 0.337 & - \\
        & 100 & 0.310 & 0.640 \\
        \hline
        $\mathbf{T}$ \textbf{ºC} & 30 & 0.370 & 0.640 \\
        \ce{CO2} = 10 vol\% & 50 & 0.212 & 0.366 \\
        $Q_T$ = 50 mL min$^{-1}$ & 75 & 0.118 & 0.211 \\
        \hline
        \textbf{\ce{CO2} (vol\%)} & 10 & 0.370 & 0.640 \\
        $Q_t$ = 50 mL min$^{-1}$ & 15 & 0.463 & 0.847 \\
        $T$ = 30 ºC & 20 & 0.545 & 1.029 \\
        & 25 & 0.631 & 1.151 \\
        \hline
    \end{tabular}
    \parbox{0.8\linewidth}{\small * C.N.: 0 ºC, 1 atm}
    \label{tab:ads_capacities_conditions}
\end{table*}

More importantly, the biomass-derived activated carbon (MSP700-A900\ce{CO2}) exhibited higher \ce{CO2} adsorption capacities than the commercial sample (WS-480) under all tested conditions. This result is particularly relevant considering that WS-480 presents an apparent higher total microporosity according to \ce{N2} adsorption analysis. Therefore, the superior performance of MSP700-A900\ce{CO2} cannot be directly attributed to the overall micropore volume. Instead, this behavior can be explained by the differences in pore size distribution and surface chemistry identified in Section \ref{sec:charac_ac}. In particular, the higher \ce{CO2} uptake of MSP700-A900\ce{CO2} is consistent with a greater contribution of ultramicropores, which are known to enhance \ce{CO2} adsorption at low partial pressures due to the stronger confinement effects within narrow pores.\cite{sevilla_sustainable_2011, presser_effect_2011} 

In addition, the broader distribution of oxygen-containing surface functionalities identified by TPD-MS suggests a more heterogeneous surface energy landscape, which may favor progressive \ce{CO2} uptake along the fixed bed under non-equilibrium conditions. Moreover, despite its higher ash content, MSP700-A900\ce{CO2} does not exhibit a reduction in adsorption capacity, indicating that the inorganic fraction does not significantly hinder the accessibility of the active microporous structure. Overall, these results demonstrate that \ce{CO2} adsorption performance is primarily governed by the distribution of ultramicropores and the surface chemistry of the material, rather than by the total porosity derived from \ce{N2} adsorption. 

Under the most representative dry post-combustion conditions evaluated in this work (15 vol\% \ce{CO2}/\ce{N2}, 30 ºC, atmospheric pressure and fixed-bed operation), the pelletized biomass-derived MSP700-A900\ce{CO2} reached a \ce{CO2} uptake of 0.847 mmol g$^{-1}$. This value is competitive with benchmark biomass-based activated carbons reported under low \ce{CO2} partial pressure, for which uptakes of 0.6–1.1 mmol g$^{-1}$ at 15 kPa and 25–50 ºC, together with fixed-bed separation of 14\% \ce{CO2}/\ce{N2} at 50 ºC, have been reported.\cite{gonzalez_sustainable_2013} More importantly, recent reviews emphasize that post-combustion adsorbent performance should be assessed under dynamic \ce{CO2}/\ce{N2} operation, rather than relying only on equilibrium uptake under pure \ce{CO2}.\cite{raganati_adsorption_2021, ibrahim_review_2025, sun_recent_2025} Although the present experiments were performed under dry conditions and therefore do not include \ce{H2O}, \ce{O2} or trace flue-gas impurities, the use of a renewable pelletized carbon in fixed-bed operation addresses key practical gaps highlighted for biomass-derived porous carbons, particularly shaping/pelletization and dynamic validation under post-combustion-relevant \ce{CO2} concentrations.\cite{sun_recent_2025, wen_recent_2024}

\section{Conclusions}
\label{sec:conclusions}
In this work, the adsorption and separation performance of a commercial coal-derived activated carbon (WS-480) and a biomass-based carbon (MSP700-A900\ce{CO2}) was evaluated by combining experimental measurements and simulations. The renewable carbon was produced via physical activation with \ce{CO2} at 900 ºC using a biochar precursor obtained from the pyrolysis of miscanthus straw pellets. \ce{N2} adsorption–desorption experiments at -196.15 ºC indicated that WS-480 exhibits an apparent higher overall degree of porosity. However, pore size distribution analysis revealed that MSP700-A900\ce{CO2} contains a larger fraction of ultramicropores ($<$0.7 nm), which resulted in higher \ce{CO2} adsorption at 0 ºC. The surface chemistry of both activated carbons was also analyzed using FTIR and TPD-MS, indicating the presence of oxygen-containing functional groups in both WS-480 and MSP700-A900\ce{CO2}. For the MSP700-A900\ce{CO2} sample, these surface groups exhibited a broader and more heterogeneous distribution over a wider range of thermal stabilities. For the computational modeling, using the CS1000a and Bhatia models as a representation of WS-480 and MSP700-A900\ce{CO2}, respectively. By subsequently introducing functional groups into these models, we accurately reproduced experimental isotherms. The results reveal clear differences between the adsorption behavior of \ce{CO2} and \ce{N2}, with all models showing a stronger affinity and higher uptake for \ce{CO2}. Comparing model types, the Bhatia-models consistently predict higher \ce{CO2} loadings and heats of adsorption relative to the CS1000a models. Furthermore, the introduction of functional groups has a pronounced effect of enhancing the \ce{CO2} uptake and raising the predicted adsorbed molar fraction of \ce{CO2}. These trends are confirmed by numerical simulations based on IAST of a 10:90 \ce{CO2}/\ce{N2} mixture, which show systematically higher \ce{CO2} loadings for both the Bhatia models and more highly functionalized structures. The GCMC simulations were also used as input for fixed-bed numerical simulations to predict breakthrough behavior, which were combined with experimental breakthrough measurements. The effects of flow rate, temperature, and \ce{CO2} concentration were systematically varied to assess dynamic adsorption performance under practical operating conditions. The modeling approach was able to accurately reproduce the experimental trends and, in most cases, predict retention times with high accuracy. Breakthrough experiments further demonstrated that the renewable carbon MSP700-A900\ce{CO2} consistently exhibited higher \ce{CO2} adsorption capacities than WS-480 under all investigated conditions. Overall, the results demonstrate that the biomass-derived carbon MSP700-A900\ce{CO2} outperforms the fossil-based commercial sample in terms of \ce{CO2} adsorption capacity, primarily due to its higher fraction of ultramicropores and its broader distribution of oxygen-containing functional groups. Furthermore, this work shows that screening representative carbon models and incorporating surface functionalization provides an efficient and reliable strategy for modeling adsorption and separation behavior in activated carbons with unknown structures.


\section*{Supporting Information}
Additional details of the experimental and simulation setup. Figure of the heats of adsorption. The data that support the findings of this study are openly available in Zenodo at http://doi.org/10.5281/zenodo.20507037.

\section*{Acknowledgements}
This work was funded by the European Union's Horizon Europe as a part of SUPERVAL project (The SUstainable Photo-ElectRochemical VALorization of flue gases) grant agreement no. 101115456. 

\section*{Author contributions}
\textbf{S. Gooijer} Investigation, Data curation, Visualization, Writing - original draft. \textbf{P. Goosen} Investigation, Methodology \textbf{C. G. Díaz-Maroto} Investigation, Data curation, Writing - review \& editing \textbf{I. Moreno} Investigation, Writing - review \& editing \textbf{S. Calero} Writing - review \& editing, Resources, Funding acquisition, Supervision. \textbf{J. Fermoso} Investigation, Data curation, Writing - original draft - review \& editing, Resources, Supervision. \textbf{J.M. Vicent-Luna} Conceptualization, Formal analysis, Writing - review \& editing, Supervision.  

\section*{Competing interests}
The authors declare that they have no known competing financial interests or personal relationships that could have appeared to influence the work reported in this paper.

\bibliography{bibliography}





\newpage

\renewcommand{\figurename}{Figure.}
\renewcommand{\thetable}{S\arabic{table}}  
\renewcommand{\thefigure}{S\arabic{figure}} 
\setcounter{figure}{0}
\setcounter{table}{0}

\onecolumngrid 

\begin{center}
    \textbf{\Huge{Supporting Information}}

    \vspace{1.0cm}

    for

    \vspace{1.0cm}

    \textbf{\Large{Influence of Ultramicroporosity and Surface Chemistry on Dynamic \ce{CO2} Capture in Activated Carbons}}
\end{center}

\section{Experimental details}
\begin{figure}[h!]
    \centering
    \includegraphics[width=0.7\textwidth]{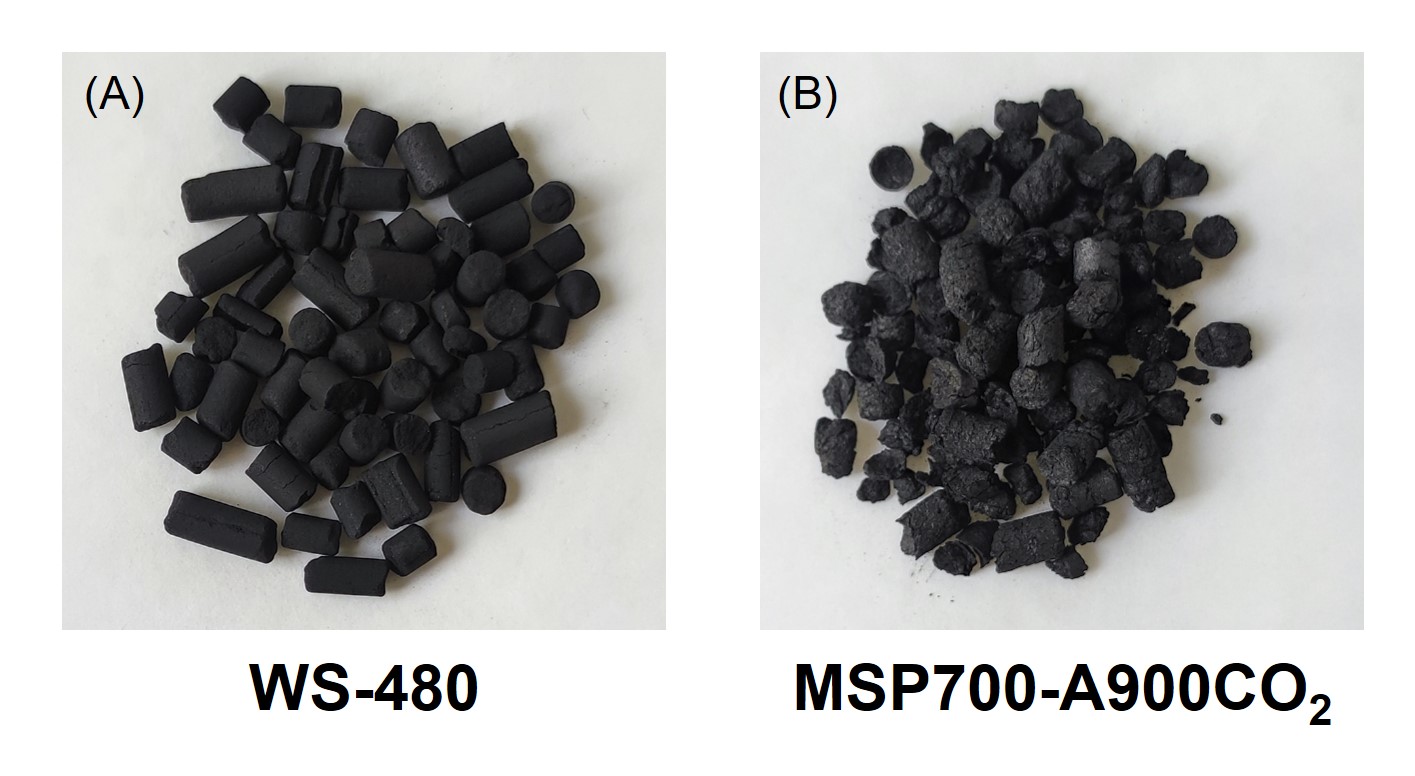}
    \caption{Images of activated carbon samples: (A) commercial derived from mineral coal high-temperature steam activated and (B) in-house \ce{CO2} activated at 900 ºC for 90 min from biochar.}
    \label{fig:S1}
\end{figure}

\begin{figure}[t]
    \centering
    \includegraphics[width=0.7\textwidth]{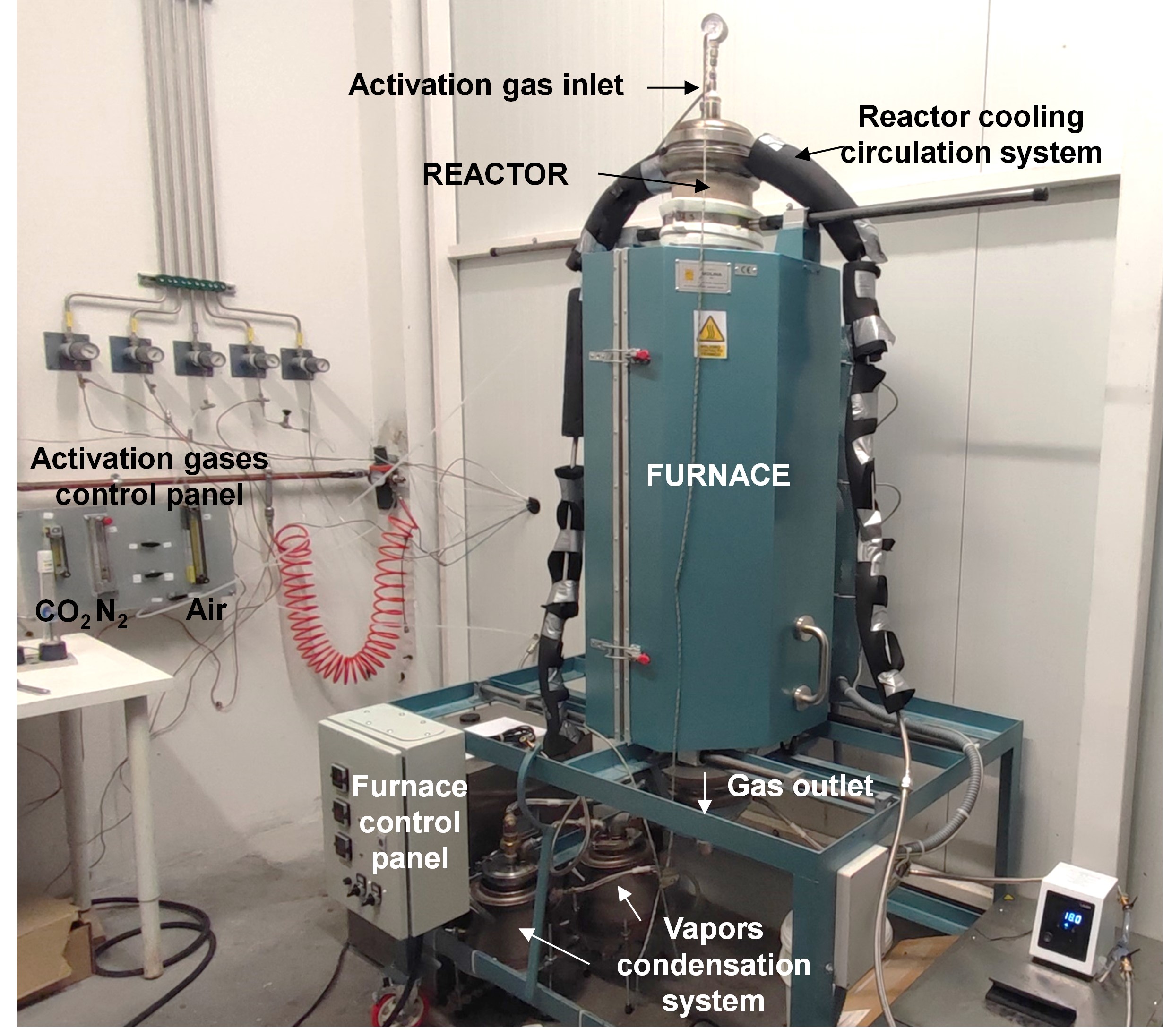}
    \caption{Picture of the pilot-plant \ce{CO2} activation plant.}
    \label{fig:S2}
\end{figure}

\begin{figure}[t]
    \centering
    \includegraphics[width=0.7\textwidth]{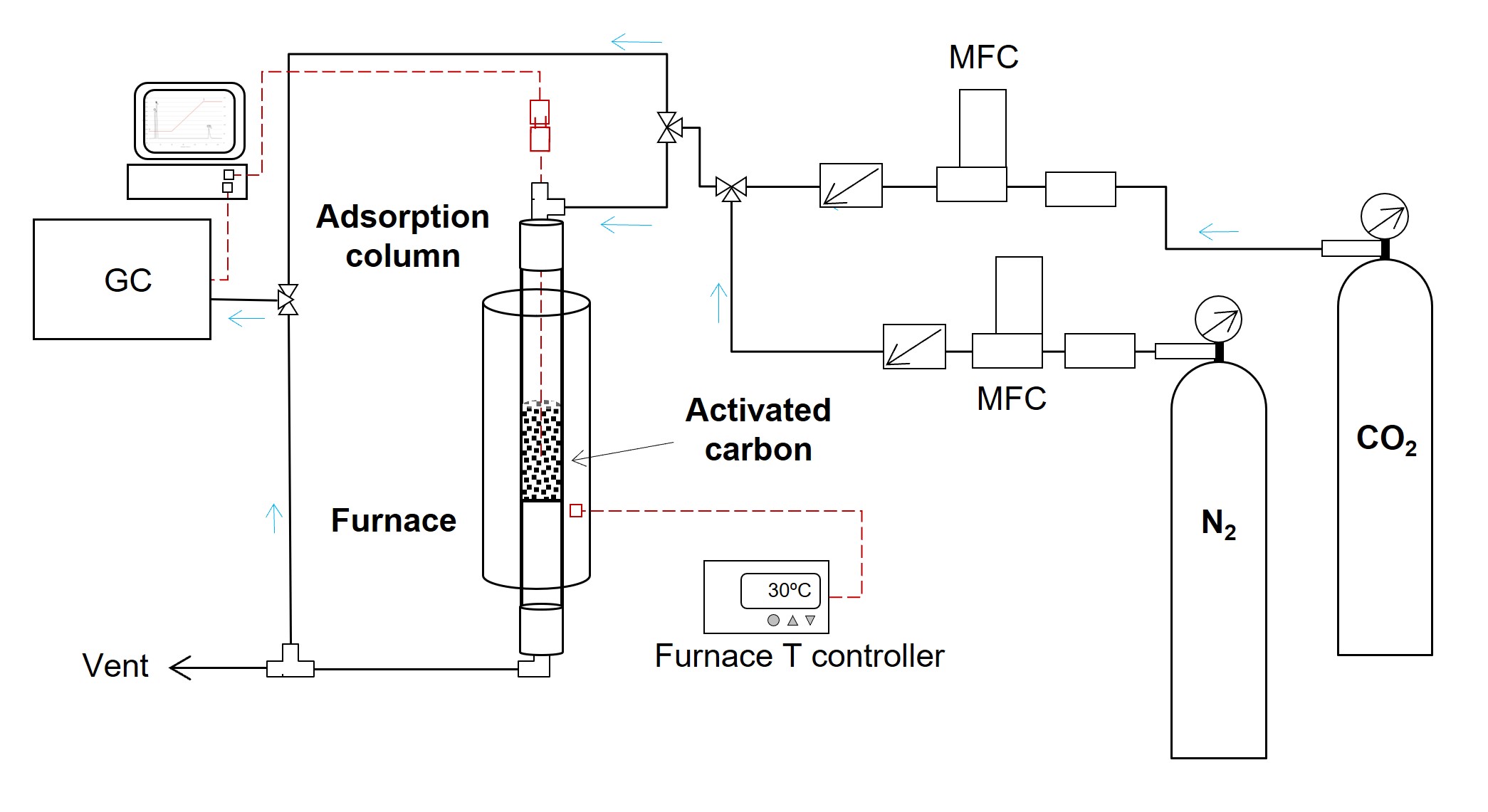}
    \caption{Schematic diagram of the lab-scale fixed-bed experimental setup for \ce{CO2} adsorption. }
    \label{fig:S3}
\end{figure}

\newpage\section{Simulation details}
Input files, structure cif files and data that support the findings of the study are openly available in Zenodo at: http://doi.org/10.5281/zenodo.20507037. 

\subsection{Activated Carbon models}
\label{sec:AC_models_info}
We used two existing activated carbon models as representations for WS-480 and MSP700-A900\ce{CO2}, respectively. For WS-480, the reconstructed model of CS1000a by Jain et al. was used.\cite{jain_molecular_2006} For MSP700-A900\ce{CO2}, we used the Bhatia 001 model from Nguyen et al.\cite{nguyen_new_2008} The models have quite different unit cell dimensions, with CS1000a having a unit cell of 50 x 50 x 50 {\AA} and Bhatia having a unit cell of 29.5 x 29.8 x 30.2 {\AA}. We tested the addition of different amounts of carboxyl and hydroxyl groups to both structures. In terms of molar fractions, we added 3 and 9\% functional groups. In terms of the number of groups per unit surface area, we tested 1.5-2 and 4.3-4.5 $\mathrm{\mu}$mol $\mathrm{m^{-2}}$. In previous work, a certain number of carboxyl and hydroxyl groups were added to the CS1000a framework to match the experimental O/C and H/C ratio of a CS1000a sample.\cite{madero-castro_adsorption_2022, jain_molecular_2006} To achieve these ratios, the functional groups were added in a 1:2 carboxyl:hydroxyl ratio. Although we added a higher concentration of functional groups in the present work, the same 1:2 carboxyl:hydroxyl ratio was maintained. In the end, the models with 9 \% (MF) and 1.5–2 $\mathrm{\mu}$mol $\mathrm{m^{-2}}$ (SA) showed the best agreement with the experimental isotherms and are therefore reported. The exact amount of functional groups added to both models are listed in Table \ref{tab:no_func_groups}.

\begin{table}[h!]
    \caption{Absolute number of framework carbon atoms, carboxyl groups, and hydroxyl groups added to the CS1000a and Bhatia models.}
    \centering
    \begin{tabular}{lccc}
        \hline
        \textbf{AC model} & \textbf{framework} & \textbf{carboxyl} & \textbf{hydroxyl} \\
        \hline
        CS1000a MF & 3785 & 129 & 258 \\
        CS1000a SA & 4090 & 84 & 168 \\
        Bhatia MF & 968 & 33 & 66 \\
        Bhatia SA & 1070 & 17 & 34 \\
        \hline
    \end{tabular}
    \label{tab:no_func_groups}
\end{table}

\subsection{GCMC simulations}
Grand canonical Monte Carlo (GCMC) simulations were performed using the RASPA3 software.\cite{ran_raspa3_2024} 10000 initialization and 50000 production cycles were performed. Each cycle $N$ number of trial moves are performed with $N$ being the number of molecules. The attempted moves are rotation, translation, and swap moves, each with equal probability. The fugacity coefficient was set to 1.0. The average properties were printed every 2000 cycles. During the simulations, radial distribution functions of the adsorbates with the framework atoms were computed. The heats of adsorption were computed using a fluctuation formula.\cite{vlugt_computing_2008} Pore size distributions (PSD) of the CS1000a and Bhatia models without added functional groups were computed using the RASPA software.\cite{dubbeldam_raspa_2016} Sigma was used as the probe distance and the maximum range of the PSD was 12 {\AA}.

Both the frameworks and adsorbates were treated as rigid, so only electrostatic and van der Waals interactions were considered. The Ewald summation method was used to calculate the electrostatic interactions.\cite{darden_particle_1993} The van der Waals interactions were treated with shifted Lennard-Jones potentials with a cutoff of 12 {\AA} without tail-corrections. To compute the cross-interactions, Lorentz-Berthelot mixing rules were applied.\cite{lorentz_ueber_1881, berthelot_sur_1898}

Point charges and Lennard-Jones parameters were taken from existing models. For the framework, the model of Jain et al. was used.\cite{jain_molecular_2006} The parameters for the carboxyl and hydroxyl groups were taken from the work of Jorge et al.\cite{jorge_simulation_2002} For \ce{CO2} and \ce{N2} we used parameters that reproduce vapor-liquid equilibrium curves.\cite{martin-calvo_effect_2011, garcia-sanchez_transferable_2009} All parameters are listed in Table \ref{tab:ff_parameters}.

\begin{table}[h!]
    \caption{Lennard-Jones parameters and partial charges of the framework and adsorbates.}
    \centering
    \begin{tabular}{lccr}
        \hline
        \textbf{atom} & $\boldsymbol{\epsilon}\mathbf{/k_B}$ \textbf{(K)} & $\boldsymbol{\sigma}$ \textbf{({\AA})}& $\mathbf{q}$ $\mathbf{(e^-)}$ \\
        \hline
        C (framework)\cite{jain_molecular_2006} & 28.00 & 3.36 & 0.000 \\
        H3 (framework)\cite{jain_molecular_2006} & 15.08 & 2.42 & 0.000 \\
        C3 (carboxyl)\cite{jorge_simulation_2002} & 0.01 & 0.50 & 0.080 \\
        C1 (carboxyl)\cite{jorge_simulation_2002} & 52.00 & 3.75 & 0.550 \\
        O1 (carboxyl)\cite{jorge_simulation_2002} & 105.7 & 2.96 & -0.500 \\
        O2 (carboxyl)\cite{jorge_simulation_2002} & 85.60 & 3.00 & -0.580 \\
        H1 (carboxyl)\cite{jorge_simulation_2002} & 0.01 & 0.50 & 0.450 \\
        C4 (hydroxyl)\cite{jorge_simulation_2002} & 0.01 & 0.50 & 0.200 \\
        O3 (hydroxyl)\cite{jorge_simulation_2002} & 78.20 & 3.07 & -0.640 \\
        H2 (hydroxyl)\cite{jorge_simulation_2002} & 0.01 & 0.50 & 0.440 \\
        C (\ce{CO2})\cite{garcia-sanchez_transferable_2009} & 29.93 & 2.745 & 0.651 \\
        O (\ce{CO2})\cite{garcia-sanchez_transferable_2009} & 85.67 & 3.017 & -0.325 \\
        N (\ce{N2})\cite{martin-calvo_effect_2011} & 38.30 & 3.306 & -0.405 \\
        com (\ce{N2})\cite{martin-calvo_effect_2011} & - & - & 0.810 \\
        \hline
    \end{tabular}
    \label{tab:ff_parameters}
\end{table}

\newpage\subsection{Adsorption isotherm fitting}

\begin{table}[h!]
    \centering
    \caption{Parameters of fit to the Sips equation (Equation 4) of simulated isotherms of \ce{CO2}.}    
    \begin{tabular}{c|ccc|ccc}
         \hline
         \multirow{2}{*} & \multicolumn{3}{|c|}{\textbf{25 ºC}} & \multicolumn{3}{|c}{\textbf{30 ºC}} \\
         & $q_{sat}$ & $b$ & $\nu$ & $q_{sat}$ & $b$ & $\nu$ \\
         \hline
         \hline
         \multirow{2}{*}{CS1000a MF} & 2.127 & $1.286\cdot10^{-05}$ & 1.019 & 1.340 & $4.050\cdot10^{-06}$ & 0.281 \\
         & 4.026 & $3.443\cdot10^{-06}$ & 0.522 & 4.001 & $5.592\cdot10^{-06}$ & 0.984 \\
         \hline
         \multirow{2}{*}{CS1000a SA} & 16.92 & $7.372\cdot10^{-07}$ & 1.035 & 1.811 & $1.340\cdot10^{-05}$ & 0.892 \\
         & 0.249 & $4.134\cdot10^{-05}$ & 0.994 & 2.835 & $3.691\cdot10^{-06}$ & 0.427 \\    
         \hline
         \multirow{2}{*}{Bhatia MF} & 3.283 & $1.853\cdot10^{-05}$ & 0.867 & 35.51 & $1.298\cdot10^{-07}$ & 1.368 \\
         & 2.431 & $4.359\cdot10^{-06}$ & 0.349 & 1.433 & $1.167\cdot10^{-05}$ & 0.952 \\  
         \hline
         \multirow{2}{*}{Bhatia SA} & 2.528 & $1.822\cdot10^{-05}$ & 0.908 & 3.213 & $3.732\cdot10^{-06}$ & 0.455 \\
         & 3.678 & $4.024\cdot10^{-06}$ & 0.479 & 2.567 & $1.578\cdot10^{-05}$ & 0.888 \\  
         \hline
         \multirow{2}{*} & \multicolumn{3}{|c|}{\textbf{50 ºC}} & \multicolumn{3}{|c}{\textbf{75 ºC}} \\
         & $q_{sat}$ & $b$ & $\nu$ & $q_{sat}$ & $b$ & $\nu$ \\
         \hline
         \hline
         \multirow{2}{*}{CS1000a MF} & 2.008 & $7.691\cdot10^{-06}$ & 0.918 & 2.535 & $2.435\cdot10^{-06}$ & 0.586 \\
         & 1.455 & $3.819\cdot10^{-06}$ & 0.334 & 0.775 & $8.070\cdot10^{-06}$ & 1.108 \\
         \hline
         \multirow{2}{*}{CS1000a SA} & 2.381 & $4.630\cdot10^{-06}$ & 0.477 & 2.173 & $2.092\cdot10^{-06}$ & 0.521 \\
         & 0.485 & $3.997\cdot10^{-05}$ & 0.595 & 1.146 & $5.598\cdot10^{-06}$ & 1.001 \\    
         \hline
         \multirow{2}{*}{Bhatia MF} & 1.988 & $3.134\cdot10^{-06}$ & 0.398 & 1.287 & $1.299\cdot10^{-05}$ & 0.834 \\
         & 2.736 & $9.735\cdot10^{-06}$ & 0.942 & 1.406 & $4.166\cdot10^{-06}$ & 0.398 \\  
         \hline
         \multirow{2}{*}{Bhatia SA} & 1.486 & $1.775\cdot10^{-05}$ & 0.784 & 12.26 & $2.520\cdot10^{-07}$ & 0.998 \\
         & 2.638 & $3.753\cdot10^{-06}$ & 0.445 & 1.894 & $3.510\cdot10^{-06}$ & 0.979 \\  
         \hline
    \end{tabular}
    \label{tab:isotherm_fit_parameters_CO2}
\end{table}

\begin{table}[h!]
    \centering
    \caption{Parameters of fit to the Sips equation (Equation 4) of simulated isotherms of \ce{N2}.}    
    \begin{tabular}{c|ccc|ccc}
         \hline
         \multirow{2}{*} & \multicolumn{3}{|c|}{\textbf{25 ºC}} & \multicolumn{3}{|c}{\textbf{30 ºC}} \\
         & $q_{sat}$ & $b$ & $\nu$ & $q_{sat}$ & $b$ & $\nu$ \\
         \hline
         \hline
         \multirow{2}{*}{CS1000a MF} & 0.004 & $1.171\cdot10^{-04}$ & 0.690 & 0.501 & $2.630\cdot10^{-06}$ & 0.527 \\
         & 1.776 & $1.009\cdot10^{-06}$ & 0.919 & 0.269 & $5.410\cdot10^{-06}$ & 0.969 \\
         \hline
         \multirow{2}{*}{CS1000a SA} & 2.815 & $4.958\cdot10^{-07}$ & 0.615 & 0.609 & $2.234\cdot10^{-06}$ & 1.021 \\
         & 0.847 & $1.915\cdot10^{-06}$ & 0.986 & 0.437 & $2.360\cdot10^{-06}$ & 0.551 \\    
         \hline
         \multirow{2}{*}{Bhatia MF} & 0.427 & $4.193\cdot10^{-06}$ & 0.297 & 0.673 & $2.767\cdot10^{-06}$ & 0.505 \\
         & 2.212 & $1.710\cdot10^{-05}$ & 0.570 & 0.236 & $9.178\cdot10^{-06}$ & 0.907 \\  
         \hline
         \multirow{2}{*}{Bhatia SA} & 0.392 & $9.082\cdot10^{-06}$ & 0.815 & 1.683 & $1.531\cdot10^{-06}$ & 0.678 \\
         & 0.405 & $4.132\cdot10^{-06}$ & 0.317 & 0.175 & $1.148\cdot10^{-05}$ & 0.968 \\  
         \hline
         \multirow{2}{*} & \multicolumn{3}{|c|}{\textbf{50 ºC}} & \multicolumn{3}{|c}{\textbf{75 ºC}} \\
         & $q_{sat}$ & $b$ & $\nu$ & $q_{sat}$ & $b$ & $\nu$ \\
         \hline
         \hline
         \multirow{2}{*}{CS1000a MF} & 0.348 & $2.977\cdot10^{-06}$ & 0.460 & 0.098 & $1.973\cdot10^{-06}$ & 0.210 \\
         & 0.202 & $6.327\cdot10^{-06}$ & 0.925 & 1.683 & $5.133\cdot10^{-07}$ & 0.973 \\
         \hline
         \multirow{2}{*}{CS1000a SA} & 0.101 & $1.043\cdot10^{-05}$ & 0.978 & 0.057 & $2.025\cdot10^{-05}$ & 0.860 \\
         & 0.824 & $1.853\cdot10^{-06}$ & 0.656 & 0.396 & $3.191\cdot10^{-06}$ & 0.544 \\    
         \hline
         \multirow{2}{*}{Bhatia MF} & 0.738 & $9.500\cdot10^{-07}$ & 1.118 & 0.230 & $4.393\cdot10^{-06}$ & 0.955 \\
         & 0.615 & $1.660\cdot10^{-06}$ & 0.679 & 0.504 & $1.812\cdot10^{-06}$ & 0.533 \\  
         \hline
         \multirow{2}{*}{Bhatia SA} & 1.477 & $1.351\cdot10^{-06}$ & 0.826 & 0.215 & $7.542\cdot10^{-06}$ & 0.853 \\
         & 0.094 & $1.601\cdot10^{-06}$ & 1.766 & 0.271 & $3.457\cdot10^{-06}$ & 0.368 \\  
         \hline
    \end{tabular}
    \label{tab:isotherm_fit_parameters_298_303_N2}
\end{table}

\newpage\subsection{Breakthrough simulations}
\label{sec:breakthrough_info}
Fixed-bed breakthrough simulations were performed using the RUPTURA software.\cite{sharma_ruptura_2023} The fitted Sips parameters listed in Tables \ref{tab:isotherm_fit_parameters_CO2} and \ref{tab:isotherm_fit_parameters_298_303_N2} of \ce{CO2} and \ce{N2}, respectively, were used. The axial dispersion coefficient was set to 0 and the mass transfer coefficient was 0.06 s$^{-1}$. The standard column conditions were 30 ºC, 50 mL min$^{-1}$, 10\% of \ce{CO2}, and 100 kPa. The temperature, flow rates, and \ce{CO2} concentrations were varied. For the Bhatia models a column length of 14 cm was used and for the CS1000a a column length of 11 cm. When the activated carbon models were functionalized, the particle density of the models changed, and as a result the column void fraction will change. The particle densities and column void fractions of all six models are listed in Table \ref{tab:column_parameters}.

\begin{table}[h!]
    \caption{Particle densities and column void fractions of the functionalized activated carbon models used as input in the fixed-bed breakthrough simulations.}
    \centering
    \begin{tabular}{lccc}
        \hline
        \textbf{AC model} & \textbf{functional groups} & \textbf{particle density} & \textbf{column void fraction} \\
        \hline
        Bhatia & no func & 875.1 & 0.594 \\
        Bhatia & MF & 964.8 & 0.632 \\
        Bhatia & SA & 926.2 & 0.616 \\
        CS1000a & no func & 727.2 & 0.378 \\
        CS1000a & MF & 803.2 & 0.437 \\
        CS1000a & SA & 784.5 & 0.424 \\
        \hline
    \end{tabular}
    \label{tab:column_parameters}
\end{table}

\newpage\section{Heats of adsorption}

\begin{figure}[h!]
    \centering
    \includegraphics[]{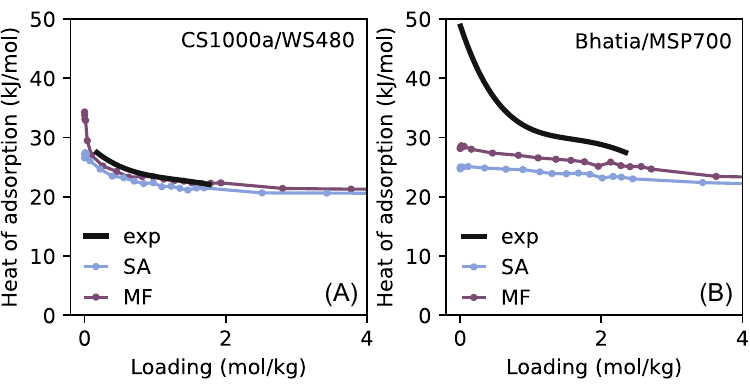}
    \caption{Simulated heats of adsorption of the SA and MF models of CS1000a (A) and Bhatia (B) plotted together with the heats of adsorption calculated from the experimental isotherms of WS-480 (A) and MSP700-A900\ce{CO2} (B) using the virial fit method at 25 ºC.}
    \label{}
\end{figure}

\end{document}